\documentclass[3p]{elsarticle}
\usepackage{lineno}

\usepackage{bm}
\usepackage{amsmath}
\usepackage{amssymb}
\usepackage{caption}
\usepackage{subcaption}
\usepackage{algorithm}
\usepackage{algpseudocode}
\usepackage{xparse}
\usepackage{bibentry}
\usepackage{makecell}
\usepackage{mathtools}
\usepackage{soul}
\algnewcommand\algorithmicinput{\textbf{INPUT:}}
\algnewcommand\INPUT{\item[\algorithmicinput]}
\algnewcommand{\LeftComment}[1]{\State \(\#\) #1}
\usepackage[acronym, toc]{glossaries}

\mathtoolsset{showonlyrefs,showmanualtags}
\newacronym{aic}{AIC}{Akaike Information Criterion}
\newacronym{bptt}{BPTT}{Backpropagation Through Time}
\newacronym{crf}{CRF}{Conditional Markov Field}
\newacronym{cnn}{CNN}{Convolutional Neural Network}
\newacronym{dfbm}{dfBm}{discrete fractional Brownian motion}
\newacronym{dfgn}{dfGn}{discrete fractional Gaussian noise}
\newacronym{cran}{CRAN}{Comprehensive R Archive Network}
\newacronym{cgsa}{CGSA}{Coarse Graining Spectral Analysis}
\newacronym{do}{DO}{Dansgaard-Oeschger} 
\newacronym{ecg}{ECG}{electrocardiogram}
\newacronym{em}{EM}{Expectation Maximization}
\newacronym{elu}{ELU}{Exponential Linear Unit}
\newacronym{fbm}{fBm}{fractional Brownian motion}
\newacronym{fgn}{fGn}{fractional Gaussian noise}
\newacronym{gem}{GEM}{Generalized Expectation Maximization}
\newacronym{gru}{GRU}{Gated Recurrent Unit}
\newacronym[plural=GP-SSMs,firstplural=Gaussian Process State Space Models (GP-SSMs)]{gpssm}{GP-SSM}{Gaussian Process State Space Model}
\newacronym[plural=GPs,firstplural=Gaussian Processes (GPs)]{gp}{GP}{Gaussian Process}
\newacronym{gps}{GPs}{Gaussian Processes} 
\newacronym{gsl}{GSL}{GNU Scientific Library}
\newacronym{hr}{HR}{Heart Rate} 
\newacronym{hmm}{HMM}{Hidden Markov Model}
\newacronym{hrv}{HRV}{Heart Rate Variability}
\newacronym{iid}{iid}{independent and identically distributed}
\newacronym{kbr}{KBR}{Kernel Based Regression}
\newacronym{mad}{MAD}{Median Absolute Deviation}
\newacronym{mlp}{MLP}{Multilayer Perceptron}
\newacronym{nca}{NCA}{Neighborhood Component Analysis}
\newacronym{nls}{NLS}{Nonlinear Least Squares} 
\newacronym{lstm}{LSTM}{Long Short-Term Memory}
\newacronym{pca}{PCA}{Principal Component Analysis}
\newacronym{psd}{PSD}{Power Spectral Density}
\newacronym{rhs}{RHS}{Right-hand side}
\newacronym{rnn}{RNN}{Recurrent Neural Network}
\newacronym{ron}{RON}{Research Octane Number} 
\newacronym{rsa}{RSA}{Respiratory Sinus Arrhythmia}
\newacronym{sde}{SDE}{Stochastic Differential Equation}
\newacronym{sdes}{SDEs}{Stochastic Differential Equations} 
\newacronym[plural=SGPs,firstplural=Sparse Gaussian Processes]{sgp}{SGP}{Sparse Gaussian Process}
\newacronym{slds}{SLDS}{Switching Linear Dynamical System}
\newacronym[plural=SSMs, firstplural=State Space Models (SSMs)]{ssm}{SSM}{State Space Model}
\newacronym{svae}{SVAE}{Structured Variational Autoencoder}
\newacronym{svi}{SVI}{Stochastic Variational Inference}
\newacronym{snr}{SNR}{Signal-to-Noise Ratio}
\newacronym{tbptt}{TBPTT}{Truncated Backpropagation Through Time}
\newacronym{osah}{OSAH}{Obstructive Sleep Apnea-Hypopnea}
\newacronym{vae}{VAE}{Variational Autoencoder}
\newacronym{vi}{VI}{variational inference}
\begin{document}
\begin{frontmatter}
\title{Stochastic embeddings of dynamical phenomena through variational autoencoders}

\author[ceu,usc]{Constantino A. Garc\'{i}a\corref{corau}}
\cortext[corau]{Corresponding author}
\ead{constantino.garciama@ceu.es}

\author[usc]{Paulo F\'{e}lix}
\author[usc]{Jes\'{u}s M. Presedo}
\author[ceu]{Abraham Otero}

\address[ceu]{Universidad San Pablo CEU, 28668, Madrid, Spain}
\address[usc]{CiTIUS, Universidade de Santiago de Compostela, 15782, Santiago de Compostela, Spain}

\begin{abstract}
  System identification in scenarios where the observed number of variables is less 
  than the degrees of freedom in the dynamics is an important challenge. In this work we 
  tackle this problem by using a recognition network to increase the observed space 
  dimensionality during the reconstruction of the phase space. The phase space is 
  forced to have approximately Markovian dynamics described by a Stochastic Differential 
  Equation (SDE), which is also to be discovered. To enable robust learning from 
  stochastic data we use the Bayesian paradigm and place priors on the drift 
  and diffusion terms. To handle the complexity of learning the posteriors, a set 
  of mean field variational approximations to the true posteriors are introduced, 
  enabling efficient statistical inference. Finally, a decoder network is used to
  obtain plausible reconstructions of the experimental data. 
  The main advantage of this approach is that the resulting model is interpretable 
  within the paradigm of statistical physics. Our validation shows that this approach 
  not only recovers a state space that resembles the original one, but it is also 
  able to synthetize new time series capturing the main properties of the experimental data.
\end{abstract}

\end{frontmatter}


\section{Introduction\label{sec:introduction}}
The study of dynamical phenomena is an important aspect in a wide variety 
of disciplines, comprising both classical fields, such as astronomy, chemistry and 
fluid mechanics; and modern fields, such as econophysics, bioinformatics, 
robotics and drone control. Modeling of such complex dynamical systems often 
relies on the concept of state space or phase space. In a deterministic system,
the state space consists of all the possible states of the system, with each state 
corresponding to a unique point in the state space. The system state at time 
$t$ contains all the information that is needed to determine the future 
system states for any instant $> t$. For a system that can be mathematically modeled,
the state space is known from the dynamic equations. 
However, researchers may lack knowledge for a completely accurate 
mathematical description of many complex natural systems. An added difficulty is that, 
in some cases, relevant variables may be missing. For example, in very complex 
systems many state variables cannot be measured directly. Even for simpler dynamical 
systems, some state variables may not be measured due to limitations in the available 
instrumentation. Furthermore, what is observed in an experimental setting is not a state
space but a set of time series representing the temporal evolution of different measurable 
properties of the system. These considerations lead to the important problem of state 
space reconstruction from a set of measurements $y(t)$ (not necessarily state variables). 
A simple idea, following the principles of classical dynamics, would be to use the 
time series itself and its derivatives 
$\dot{y}(t)$, $\ddot{y}(t), \ldots$, to build a state space.
This idea is not usually feasible due to the fact that the measurement noise 
gets amplified with each derivative and hence even second order derivatives can
look like noise.

The state space reconstruction problem was theoretically solved by the 
Takens' theorem \cite{takens1981detecting}. Takens' theorem states
that if we are able to observe a single scalar quantity $y(t)$ that depends
on the current state of the system $\bm{x}(t)$,
\begin{equation}
    y(t) \triangleq y(\bm{x}(t)),
    \label{eq:vaele:y_as_fun_x}
\end{equation}
then, under quite general conditions, the structure of the multivariate phase
space can be unfolded from this set of scalar measurements $y(t)$ by means of the 
so-called delay embedding. A delay embedding in $d$ dimensions is
formed by using delay coordinates:
\begin{equation}
 \bm{y}(t) = [y(t), y(t-\tau), y(t-2\tau), \dots, y(t - (d - 1)\tau)]^ T,
 \label{eq:delay_embedding}
\end{equation}
where the time distance $\tau$ between adjacent coordinates is usually 
referred to as the time lag or time delay. Takens' theorem guarantees that
the new geometrical object formed by $\bm{y}(t)$ is topologically
equivalent to the original state space if the embedding dimension $d$ is 
sufficient large. Specifically, the delay map should use $d > 2 D_F + 1$ 
dimensions, being $D_F$ the number of the active degrees of freedom of the 
system \cite{sauer1991embedology}. According to the Takens' theorem, the value 
of the lag $\tau$ is arbitrary.  However, in practice the proper choice of 
$\tau$ is important.  If $\tau$ is too small the delay vectors will be very 
similar between them, and therefore they will tend to cluster around the 
bisectrix of the state space. On the other hand, if $\tau$ is too large, the 
delay vectors will be almost uncorrelated, resulting in a very complex phase 
space. Several methods have been proposed to select the time lag $\tau$. A 
simple yet effective strategy is to select $\tau$ as the first minimum of the 
autocorrelation function or the average mutual information 
\cite{kantz2004nonlinear}. Nowadays, there is still active research applying 
new ideas to the selection of both the embedding dimension $d$ and the time lag 
$\tau$~\cite{vlachos2010nonuniform,tran2019topological}. 

Recently, different works have successfully built time-delayed state spaces from a single
observable $y(t)$ by stacking lagged versions of $y(t)$ into a Hankel matrix and then using 
Singular Value Decomposition (SVD)~\cite{press1992numerical}. The resulting eigen-time-delay vectors connect the Takens embedding theorem with the Koopman
operator, which permits building high-dimensional, although linear, state spaces~\cite{schmid2010dynamic,brunton2017chaos}.
Extensions of this approach can even be used with Markov processes (instead of deterministic
systems)~\cite{williams2015data}. Recent works have leveraged Koopman operator theory with 
deep learning methods to develop fully data-driven but interpretable embeddings~\cite{lusch2018deep, takeishi2017learning}.

There is a trade-off between having a linear state space and the high dimensionality of such 
space. It may be argued that the learning of nonlinear state spaces may result in lower dimensionalities,
which may facilitate the interpretability of the model. 
Works such as~\cite{ayed2019learning,ouala2019learning,raissi2018hidden}
permit discovering an augmented state space and, at the same time, finding the equations
that describe this latent space using machine learning. This approach is suggestive since it is 
very natural: the phase space is discovered by only requiring that the resulting state 
space vectors have enough information to: 1) reconstruct the original observed time series
and 2) permit the accurate prediction of future states.

Despite the impressive progress in tackling the state space reconstruction problem, the 
above-mentioned techniques still have limitations. First, the Takens' theorem only guarantees 
the preservation of the attractor's topology. Hence, the geometry of the attractor is not 
necessarily preserved, since bending and stretching are allowed mappings. Therefore close points
on the original attractor may end up far in the reconstructed state space. This makes 
Takens' theorem sensible to noise, since small fluctuations could have 
large effects on the delay reconstruction~\cite{yap2014first}. 

Furthermore, although the noise may be small enough to be ignored, the
assumption that the experimental data can be described by a deterministic
differential equation is often unrealistic. Complex systems usually involve 
a number of degrees of freedom much larger than what an experiment can resolve. In this 
case, considering a dual deterministic-stochastic model in which noise 
represents the unresolved deterministic dynamics may be more appropriate 
than a pure deterministic one. In this sense, an important limitation 
of~\cite{ayed2019learning,ouala2019learning, raissi2018hidden} is that they focus on deterministic phenomena. 

The stochastic version of an equation of motion is the \gls*{sde} or Langevin equation. 
Intuitively, a \gls*{sde} couples a deterministic trend with noisy fluctuations that 
introduce uncertainty in the evolution of the system. The \gls*{sde} for a $d$-dimensional
state vector $\bm{x}(t)$ reads
\begin{equation}
    dx_i(t) = f_i\big(\bm{x}(t)\big)dt + 
    \sum_{j=1}^d \sqrt{g_{ij}\big(\bm{x}(t)\big)} dW_j(t), \qquad i=1, 2, \dots, d
    \label{eq:vaele:vector_langevin}
\end{equation}
where $\{W_j(t)\}_{j=1}^d$ denote $d$ independent Wiener processes. 
The Wiener process has independent
Gaussian increments $W(t+\tau)-W(t)$ with zero mean and variance $\tau$. Therefore, 
the set $\{W_j(t)\}_{j=1}^d$ acts as the source of randomness of the system. 

\glspl*{sde} have revealed as a valuable tool for building coarse-grained models. 
Indeed, \glspl*{sde} have already been successfully applied 
to many problems (see~\cite{friedrich2011approaching} for a review). However, experimental 
time series have a very limited number of degrees of freedom, which may not even span 
those required by the reduced-order models. Hence, Equation~\eqref{eq:vaele:vector_langevin} is also
concerned with the same issues previously discussed when analyzing experimental data.

Since a \gls*{sde} generates a Markov process, we could expect that a time
series from a single observable could be described by a Markov process of some
order, probably larger than the order of the complete \gls*{sde} process
but somewhat related, in analogy with the Takens' theorem. Unfortunately, this assumption
is wrong and no stochastic-embedding theorem exists \cite[Chapter~12]{kantz2004nonlinear}.
The reason for this is that a scalar time series originated from a continuous 
Markov process is not longer Markovian since it has infinite memory 
\cite{risken1996fokker}. However, in most cases the memory decays exponentially 
fast and hence a Markov process can effectively approximate the time series 
under study. When using a Markov process of order $n$ for analyzing the scalar 
time series $y_k = y(k \cdot \Delta T)$ the dynamics of the time series are 
determined by the transition probabilities 
$p\left(y_k \mid y_{k-1}, y_{k-2}, \dots, y_{k-n}\right)$. This may be interpreted as an embedding
where $d=n$ and $\tau = \Delta T$, although there is
no theorem that guarantees that it will have nice properties for the analysis
of the underlying dynamical system. Note that, in this context, the new state
space may be seen as the result of a transformation that is able to represent a 
scalar time series as a vector Markov process. Having such a transformation will 
enable the analysis of non-Markovian time series with Markovian theory, for which a 
large body of research exists.

Hence, an interesting extension of~\cite{ayed2019learning,ouala2019learning,raissi2018hidden} would 
be allowing the reconstruction of the state space of stochastic systems while learning 
the equations. However, robust learning of the state space is hard, specially with stochastic data. For example, 
we may imagine a time series in which, driven by noise, the system visits a region of the
phase space only once. When learning from only a few samples, model-based methods usually suffer from model bias~\cite{deisenroth2011pilco}: they 
become overconfident about their own predictions. To tackle this issue, a probabilistic
modeling supported by Bayesian learning provides a more robust methodology.

Indeed, learning of non-linear and stochastic \glspl*{ssm} from data has become a key issue in
both control systems and reinforcement learning, resulting in a large body 
of research in modern \glspl*{ssm}~\cite{murphy2012machine}. A typical discrete-time \gls*{ssm} reads as
\begin{align}
  &\bm{x}_{t} = \bm{f}(\bm{x}_{t-1}) + \bm{\epsilon}_{t},\\
  &\bm{y}_t = \bm{h}(\bm{x}_t) + \bm{\gamma}_t,
  \label{eq:ssm}
\end{align}
where $\bm{x}_t$ is the latent state, $\bm{y}_t$ is the observed space, 
and $\bm{\epsilon}_t$ and $\bm{\gamma}_t$ are the transition and observation noises,
respectively. 

Placing priors on both the transition and observation functions ($\bm{f}$ and $\bm{h}$) enables
the Bayesian treatment of the problem, which has proved to very effective for a robust learning. 
Since the most common prior for functions are \glspl*{gp} the resulting models are referred to 
as \glspl*{gpssm}. A \gls*{gp} assumes that any finite number of function points
$\left[f(\bm{x}_1),f(\bm{x}_2), \dots, f(\bm{x}_n) \right]^T$  have a 
joint Gaussian distribution. Therefore, a \gls*{gp} is fully specified by a mean
function $m(\bm{x})$ and a kernel (or covariance) function $k(\bm{x}, \bm{x}')$. We 
note this as $f(\bm{x}) \sim \mathcal{GP}(m(\bm{x}),k(\bm{x},\bm{x}'))$.
The properties of the kernel determine the basic behaviour of the functions 
that we want to model.

The first fully Bayesian technique that used a \gls*{gpssm} for system identification is, to 
the best of our knowledge~\cite{frigola2013}, which proposed a particle Markov Chain Monte 
Carlo for learning the latent space. An important drawback
of this approach is that it has a large computational cost. To tackle this issue, the same
authors later proposed a variational Bayesian learning algorithm~\cite{frigola2014}, which 
permitted to trade-off between computational burden and model capacity. Following these 
works, \cite{eleftheriadis2017identification} proposed the use of bidirectional \gls*{rnn}
as a recognition model to facilitate learning the transition model from the unobserved 
latent space. Intuitively, a recognition network permits obtaining probabilistic guesses 
of the latent space from the observed space, which aids in building the state space. 

The use of recognition networks is not exclusive of \gls*{gpssm}. For example, recognition 
networks were also employed in ~\cite{krishnan2015deep}, which generalized Kalman filters 
by parametrizing them with neural networks and explicitly considered \glspl*{rnn} as a
possible recognition network; and in~\cite{johnson2016composing}, which introduced \glspl*{svae}.
A \gls*{svae} uses a recognition network (also referred to as encoder) to map the input data to a latent 
space with a structured probability distribution (hence the name). Another neural network, 
named the decoder, is able to map samples from the latent space to realistic realizations 
of the observed data.

Both \glspl*{gpssm} and \glspl*{svae} can be applied to the system identification problem, in which we
learn a latent space from the observation of raw data. However, there is usually the 
implicit assumption that the observed space has a large number of degrees of freedom. 
In this case, finding a state space implies a reduction of dimensionality, which may be 
useful for better understanding the dynamical properties of the system or for efficient 
simulations.

In this paper, we give a step forward towards modeling incomplete time series, 
in the sense that the number of observed variables are smaller than the degrees of freedom 
in the dynamics. Given an incomplete time series, is it possible to disentangle the dynamics 
and reconstruct the underlying state space? Note that this implies increasing the dimensionality 
of the observed space when creating the latent space. Hence, we tackle the reconstruction problem for both deterministic and 
stochastic data following the intuitive formulation of~\cite{ayed2019learning,ouala2019learning, raissi2018hidden}.
That is, state space reconstruction is achieved by simultaneously discovering the 
equations describing the state space, and by also requiring the latent space to yield accurate 
reconstructions of the observed space. 

Section~\ref{sec:vaele:sde_svae} starts detailing the model used for state space reconstruction.
To facilitate augmenting the dimension of the observed time series to build the phase space we 
make use of a recognition network. The state space is then forced to approximately have 
Markovian dynamics described by a \gls*{sde}. For robust learning from stochastic data we adopt 
the Bayesian paradigm, placing proper priors on the drift and diffusion functions. Finally, a decoder network is used to obtain plausible reconstructions of experimental data. 
We refer to the resulting model as the \gls*{sde}-\gls*{svae}, since the structured latent space
is an stochastic process modeled by an~\gls*{sde}. However, it can also be interpreted as 
a Bayesian \gls*{gpssm} when discretizing the \gls*{sde} under the Euler-Maruyama scheme. 

As in any Bayesian framework, after setting the model our aim is to infer the distributions of its
parameters after observing the experimental data, the so-called posterior distributions.
Unfortunately, learning the posteriors from data of the \gls*{sde}-\gls*{svae} model is very challenging and hence, some 
approximations must be derived. A set of mean field variational approximations to the true 
posteriors are introduced, enabling efficient statistical inference. This is usually referred to 
as \gls*{vi}. The paper actually proceeds in two steps. Section~\ref{sec:vaele:global} discusses 
the introduction of inducing points for approximating the posterior of the drift function. 
This is key since it also permits the use of batches of experimental data during learning, resulting 
in the \gls*{svi} algorithm. The main advantage of \gls*{svi} is that it is able to handle datasets 
of arbitrary size. Section~\ref{sec:vaele:variational_approx} then proceeds to detail the variational approximations 
of the rest of posterior distributions, which finally enable efficient learning with the 
algorithm detailed in Section~\ref{sec:vaele:lower_learning}.

Section~\ref{sec:vaele:validation} contains experimental simulations over synthetic datasets 
with physical and biological motivations. Finally, Section~\ref{sec:vaele:discussion} discusses
the main results and suggests possible research directions.

\section{SDE-SVAE\label{sec:vaele:sde_svae}}
In this section we develop a \gls*{ssm} based on \glspl*{sde}. We consider a collection
of discrete-time signals obtained from sampling different realizations of 
a continuous process with 
\begin{equation*}
        \bm{y}^{(r)}_t=\bm{y}^{(r)}(t \cdot \Delta T),\qquad r=1,2,\dots,R;\qquad t=1,2,\dots,N,
\end{equation*}
being $\Delta T$ the sampling period, $R$ the number of different realizations 
of the process, and $N$ the total number of samples. To keep the notation 
uncluttered, we shall eliminate the superscript $(r)$ from the formulas, as
if $R=1$, unless necessary. Furthermore, we will note a whole time series as $\bm{y}_{1:N}$.

We assume that the observations $\bm{y}_{1:N}$ are generated from a 
$d$-dimensional latent state space $\bm{x}_{1:N}$. Furthermore, we also assume that 
this state space can be modeled as the sampled version of a process $\bm{x}(t)$ described 
by a \gls*{sde} (see Equation~\eqref{eq:vaele:vector_langevin}). To simplify the 
state space, we assume that the diffusion matrix from 
Equation~\eqref{eq:vaele:vector_langevin} is diagonal and constant, i.e.,
$g_{ii}$ does not depend on $\bm{x}(t)$:
\begin{equation}
    dx_i(t) = f_i\big(\bm{x}(t)\big)dt + \sqrt{g_{ii}}\, dW_i(t) \qquad i=1, 2, \dots, d
    \label{eq:vaele:langevin_model}
\end{equation}
This assumption is guided by the existence of a transformation that is able to leave the 
diffusion term independent of the state $\bm{x}(t)$: the so called Lamperti 
transformation~\cite{moller2010state}. The Lamperti transformation can be applied to 
\glspl*{sde} of the type
$$d\bm{x}(t) = f(\bm{x}(t))dt + \bm{G}(\bm{x}(t), t)\bm{R}(t)d\bm{W}(t),$$
where $\bm{R}(t)$ is any matrix and $\bm{G}(\bm{x}(t), t)$ is a diagonal matrix fulfilling
\begin{equation}
 \bm{G}_{ii}(\bm{x}(t), t) = G_i(x_i(t), t),
 \label{eq:lamperti_condition}
\end{equation}
which for the moment we consider rich enough to construct valuable embeddings. The 
applicability of the Lamperti transformation is limited by the fact that it is constructed 
using the diffusion term. Hence, when the diffusion term is not known, the Lamperti 
transformation cannot be applied. However, in our approach we are building a new state space 
$\bm{x}_{1:N}$ from scratch. Since the Lamperti transformation guarantees that a 
representation with a state independent diffusion exists (and assuming that 
Equation~\eqref{eq:lamperti_condition} is met), we can select this 
representation as our state space without loss of generality.

If we assume that the Euler-Maruyama discretization scheme holds~\cite{kloedenPlaten}, 
then Equation~\eqref{eq:vaele:langevin_model} becomes
\begin{equation}
    \Delta x_{i,t}= f_i\big(\bm{x}_t\big)\Delta T + \sqrt{g_{ii}} (W^i_{t+1} - W^i_t),
  \qquad i=1, 2, \dots, d
  \label{eq:vaele:euler_maruyama}
\end{equation}
where $x_{i, t}$ denotes the $i$-th dimension of the $t$-th sample and therefore 
$\Delta x_{i, t} = x_{i, t+1} - x_{i, t}$. Hence, the discrete
transition probabilities can be approximated with a single multivariate Gaussian
with diagonal covariance matrix:
\begin{equation}
    p(\bm{x}_{t+1}|\bm{x}_t, \bm{f}, \bm{g}) = 
    \mathcal{N}\left(\bm{x}_{t+1}\mid \bm{x}_{t} + \bm{f}(\bm{x}_t)\Delta T,
    \text{diag}(\bm{g})\Delta T\right),
    \label{eq:vaele:transition_probabilities}
\end{equation}
where $\bm{f}(\bm{x}) = [f_1(\bm{x}), f_2(\bm{x}), \dots, f_d(\bm{x})]^T$
and $\bm{g}=[g_{11}, g_{22}, \dots, g_{dd}]^T$.

Note that, since the state space is not observed, both $\bm{f}(\cdot)$ and $\bm{g}$
are not known and must be learned. To enable rich dynamics, we parametrize
both $\bm{f}(\cdot)$ and $\bm{g}$ with a flexible model. Following previous 
works on \gls*{sde} estimation~\cite{ruttor2013approximate,voila}, 
the drift function is modeled with a \gls*{gp}. On the other hand, 
we note that $g_{ii}$ plays the role of the variance in 
Equation~\eqref{eq:vaele:transition_probabilities}. Since the Normal-Gamma 
distribution is the conjugate prior of a Gaussian distribution with unknown mean
and precision, we propose the use of a \gls*{gp}-Gamma distribution for modeling 
the drift and diffusion terms. We note 
$$
f(\bm{x}), \lambda \mid \bm{\theta}\sim
\text{GP-Gamma}(f(\bm{x}), \lambda \mid m(\bm{x}), k(\bm{x}, \bm{x}'), 
\alpha, \beta; \bm{\theta})
$$ 
implying that
    \begin{align}
    \label{eq:vaele:gp_gamma}
        & \lambda \sim  \Gamma(\lambda\mid \text{shape}=\alpha, \text{rate}=\beta), \\
        & f(\bm{x})|\bm{\theta}, \lambda \sim\mathcal{GP}
        \bigg(m(\bm{x}), \frac{1}{\lambda}k(\bm{x}, \bm{x}',\bm{\theta})\bigg).
    \end{align}
Therefore, we model the drift and diffusion terms as 
    \begin{align}
        \label{eq:vaele:full_model}
        &g_{ii} = \frac{1}{\lambda_i}, \qquad i=1, 2, \dots, d,\\
        &f_i(\bm{x}), \lambda_i \mid \bm{\theta}_i\sim
        \text{GP-Gamma}(f_i(\bm{x}), \lambda_i \mid 0, k_i(\bm{x}, \bm{x}'), 
        \alpha_i, \beta_i; \bm{\theta}_i),
    \end{align}
where we have used a zero prior for the mean of $f_i$ for symmetry reasons.
For convenience, we shall denote 
$\bm{\lambda} = [\lambda_1, \lambda_2, \dots, \lambda_d]^T$. Also, we
let the $\Delta T$ term to be absorbed by the parameters of the 
\gls*{gp}-Gamma distribution, since 
$$f(\bm{x})\Delta T, \lambda/\Delta T \sim
\text{GP-Gamma}\big(f(\bm{x})\Delta T, \lambda/\Delta T \mid
m(\bm{x})\Delta T, k(\bm{x}, \bm{x}')\Delta T,
\alpha, \beta \Delta T; \bm{\theta}\big)$$
if 
$$f(\bm{x}), \lambda\sim 
\text{GP-Gamma}\big(
f(\bm{x}), \lambda \mid m(\bm{x}), k(\bm{x}, \bm{x}'), \alpha, \beta; 
\bm{\theta}\big).$$
Therefore, $\Delta T$ shall subsequently be ignored in our exposition.

\begin{figure}
  \centering
  \begin{subfigure}[t]{0.47\textwidth}
      \includegraphics[width=0.9\textwidth]{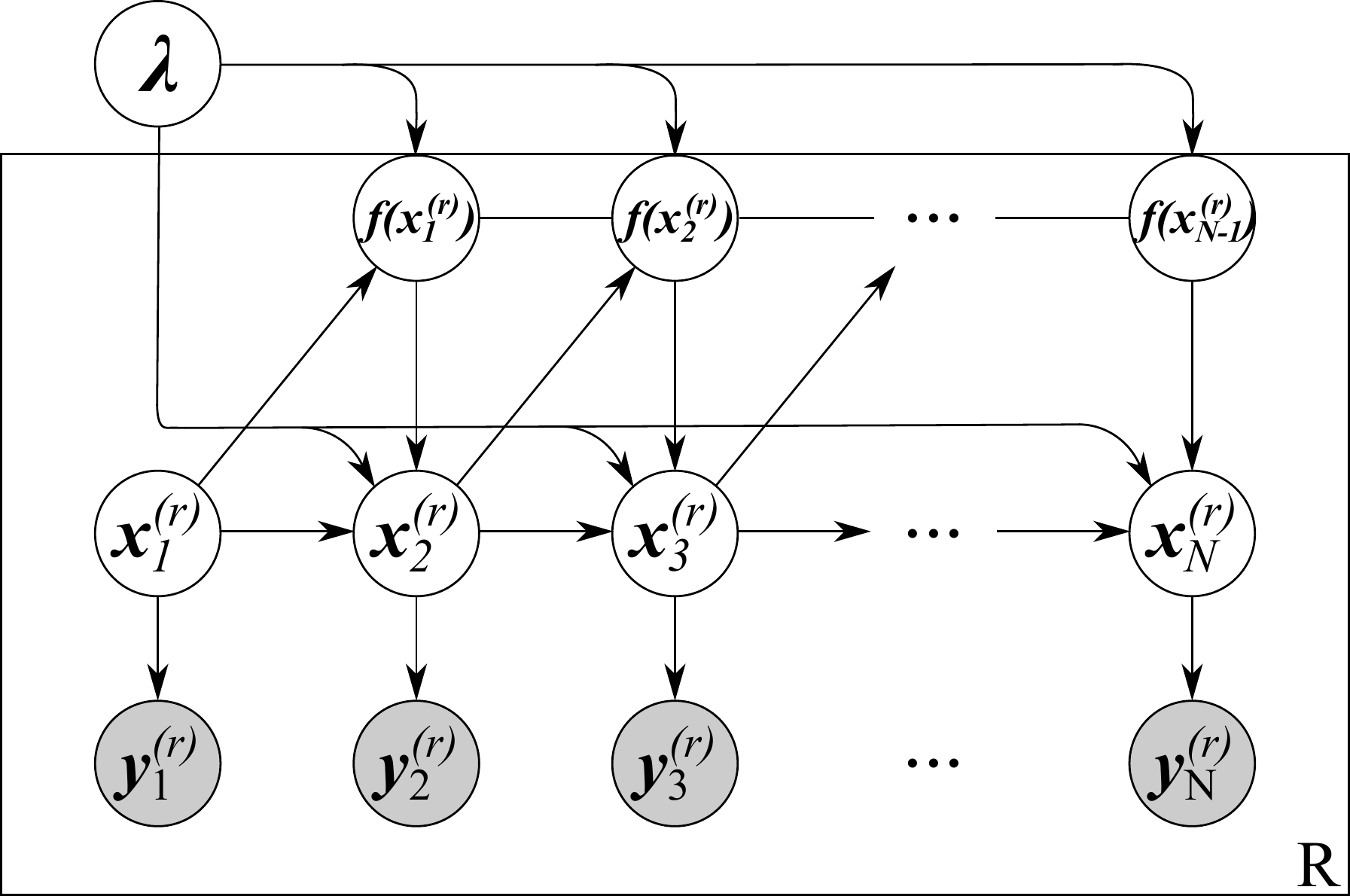}
      \caption{\label{fig:vaele:complete_graphical_model}}
  \end{subfigure}
  \begin{subfigure}[t]{0.47\textwidth}
    \includegraphics[width=0.9\textwidth]{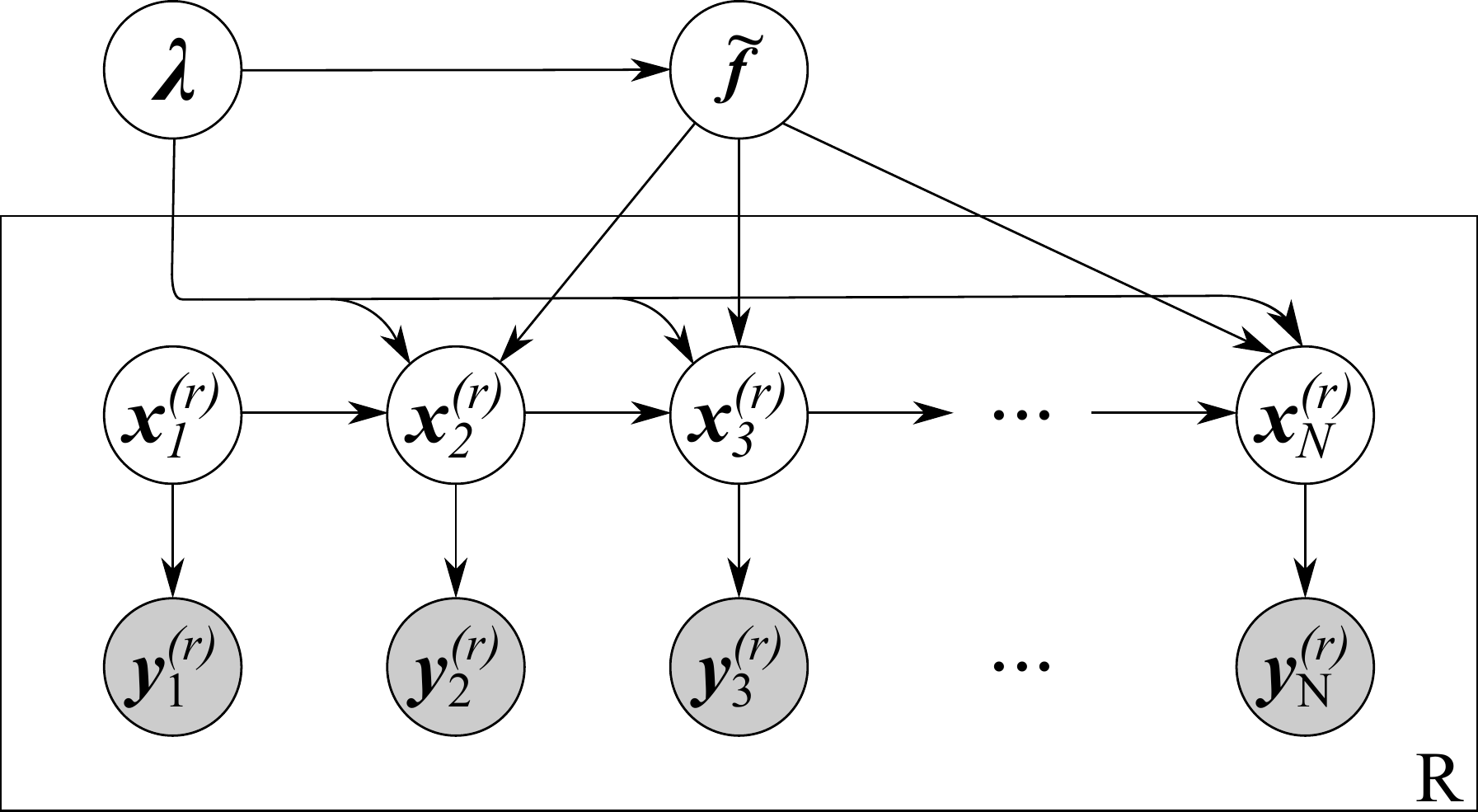}
  \caption{\label{fig:vaele:sparse_generative}}
  \end{subfigure}
  \caption{(a) Generative graphical model for a phase
    space with \gls*{sde}-dynamics. (b) Generative model with inducing points compatible
    with \gls*{svi}. Gray nodes represent observed variables and the plate notation indicates
    the repetition of random variables.
  }
\end{figure}

Each sample from the state space $\bm{x}_t$ generates an observation $\bm{y}_t$.
To avoid the problem of non-identifiability between emissions and transitions~\cite{frigola2014}
we use a simple linear emission model:
\begin{align}
        &\bm{y}_t \mid \bm{x}_t, D \sim 
        \mathcal{N}\left(\bm{y}_t \mid \bm{D}\bm{x}_t + \bm{d}, \bm{S} \right),
        \label{eq:vaele:decoder}
\end{align}
where the parameters $D=(\bm{D}, \bm{d}, \bm{S})$, have been noted with a 
mnemonic for \textit{D}ecoder. The generative model that results from these assumptions is 
shown in Figure~\ref{fig:vaele:complete_graphical_model}. It is worth noting that, under 
the Euler-Maruyama scheme, the \gls*{sde} formulation is equivalent to a \gls*{gpssm}.

As illustrated by Figure~\ref{fig:vaele:complete_graphical_model}, the \gls*{gp} introduces 
dependencies between the latent state space vectors. This is one of the main difficulties
preventing efficient \gls*{gpssm} inference schemes since, for example, even in a dataset 
with a single example, to sample $\bm{x}_T$ we would need to condition on all the previous 
$T-1$ points~\cite{ialongo2019overcoming}. Recent work has tackled the issue by introducing
novel factorized posteriors and effective sampling schemes~\cite{ialongo2019overcoming,curi2020structured}. 
In this paper, taking inspiration from~\cite{hensman2013gaussian},
we scape from the dependencies introduced by the \gls*{gp} by using a \gls*{sgp}, 
which approximate the full \gls*{gp} by using a small set of inducing points (in the same 
way we may approximate a function $f(x)$ by using a collection of points $\bm{f}$).
An additional advantage of this approach is that it enables the use of \gls*{svi}, which 
can be subsequently used to handle large datasets with multiple time series realizations.
Hence, following~\cite{hensman2013gaussian}, we shall introduce
a set of inducing points in our model to act as global variables, as required by 
\gls*{svi}. 

\subsection{Inducing points as global variables for \gls*{svae}\label{sec:vaele:global}}
The \gls*{sgp} approximation is built following~\cite{voila}. Therefore, our inducing 
variables shall be the function points that result from evaluating $\bm{f}(\bm{x})$ at $m$
pseudo-inputs 
$\bm{\tilde{X}}=\{\bm{\tilde{x}}_j:\bm{\tilde{x}}_j \in \mathbb{R}^d\}_{j = 1}^m$.
We have noted the inducing points with uppercase to highlight that they 
are a collection of points in $\mathbb{R}^d$, and because we can arrange them 
as a matrix with $m$ rows and $d$ columns. Although a more general treatment
is possible, for the sake of simplicity we assume that the $d$ \gls*{gp}s 
modeling the vector function $\bm{f}(\cdot)$ share 
the same set of pseudo-inputs. Since Equation~\eqref{eq:vaele:euler_maruyama}
decomposes the equations of motion in $d$ independent equations, we shall 
study each component of $\bm{f}(\cdot)$ independently. The $i$-th component of the 
inducing points shall be noted as 
$\bm{\tilde{f}}_i = \{f_i(\bm{\tilde{x}}_j): \bm{\tilde{x}}_j \in \bm{\tilde{X}}\big\}$,
and it is a vector of length $m$. From our previous assumptions, it follows that 
the $i$-th component of the inducing-points only depends on the values of the 
$i$-th \gls*{gp}. Furthermore, since $\bm{\tilde{f}}_i$ is derived from the same 
\gls*{gp} as $f_i(\cdot)$, the following conditional distribution holds:
\begin{equation*}
   \begin{bmatrix}
     f_i(\bm{x})\\
     \bm{\tilde{f}}_i
 \end{bmatrix}\;\vert \lambda_i
   \sim
  \mathcal{N}\left(
  \begin{bmatrix}
      0\\
       \bm{0}
     \end{bmatrix}
  ,
  \frac{1}{\lambda_i}\begin{bmatrix} 
      \bm{K}_i(\bm{x}, \bm{x})&\bm{K}_i(\bm{x}, \bm{\tilde{X}})\\
      \bm{K}_i(\bm{\tilde{X}}, \bm{x}) &\bm{K}_i(\bm{\tilde{X}}, \bm{\tilde{X}})\\
  \end{bmatrix}\right)\qquad i=1, 2, \cdots, d,
\end{equation*}
where $\bm{x} \in \mathbb{R}^d$, $\bm{K}_i(\bm{x}, \bm{x})$ is a $1\times 1$ 
matrix, and the matrices 
$\bm{K}_i(\bm{x}, \bm{\tilde{X}})$, $\bm{K}_i(\bm{\tilde{X}}, \bm{x})$ and  
$\bm{K}_i(\bm{\tilde{X}}, \bm{\tilde{X}})$
should be read as the $1 \times m$, $m \times 1$ and $m \times m$ matrices.
Note that $f_i(\cdot)$ is a scalar 
function and it should not be confused with the $i$-th dimension of the 
inducing-points $\bm{\tilde{f}}_i$. The latter is the result
of evaluating $f_i(\cdot)$ on a fixed set of points (the pseudo-inputs $\bm{\tilde{X}}$) 
and hence, it is possible to interpret it as a multivariate Gaussian random variable.
The distribution of $f_i(\bm{x})$ conditioned on $\bm{\tilde{f}}_i$ and $\lambda_i$
can be written as
\begin{align}
    &f_i(\bm{x})\mid \bm{\tilde{f}}_i, \lambda_i \sim 
    \mathcal{N}\bigg(f_i(\bm{x})\mid  \bm{A}_i(\bm{x}) \bm{\tilde{f}}_i,
    \bm{P}_i(\bm{x}) / \lambda_i \bigg)\qquad \text{with}    \\
    &\bm{A}_i(\bm{x}) = \bm{K}_i(\bm{x}, \bm{\tilde{X}})\cdot 
    \bm{K}_i^{-1}(\bm{\tilde{X}}, \bm{\tilde{X}}),\\
    &\bm{P}_i(\bm{x}) = \bm{K}_i(\bm{x}, \bm{x}) -
    \bm{K}_i(\bm{x}, \bm{\tilde{X}})\cdot 
    \bm{K}_i^{-1}(\bm{\tilde{X}}, \bm{\tilde{X}}) \cdot 
    \bm{K}_i(\bm{\tilde{X}} , \bm{x}).
   \label{eq:vaele:transformation_matrix}
\end{align}
Note that we have used a bold notation for $\bm{P}_i(\bm{x})$ because we compute
its value through multiplications of matrices. However, the final 
$\bm{P}_i(\bm{x})$ from Equation~\eqref{eq:vaele:transformation_matrix}
is a $1\times 1$ matrix and should be read as a scalar.

To ``eliminate'' the \gls*{gp} from the model shown in 
Figure~\ref{fig:vaele:complete_graphical_model}, we need to replace the role of 
the conditional distribution involving the \gls*{gp}, 
$p(\bm{x}_{1:N}\mid \bm{f}(\cdot), \bm{\lambda}, \bm{\theta})$, by the 
conditional distribution involving the inducing-points,
$p(\bm{x}_{1:N}\mid \bm{\tilde{f}}, \bm{\lambda}, \bm{\theta})$. 
To that end, we lower bound the conditional distribution by using Jensen's inequality (see, 
for example,~\cite{murphy2012machine}):
\begin{align}
  \log p(\{\bm{x}^{(r)}_{1:N}\}\mid \bm{\tilde{f}}, \bm{\lambda}, \bm{\theta}) &=   
    \log \mathbb{E}_{p(\bm{f}\mid \bm{\tilde{f}}, \bm{\lambda}, \bm{\theta})} 
    \left[p(\{\bm{x}^{(r)}_{1:N}\}\mid \bm{f}, \bm{\lambda}, \bm{\theta}) \right]\\
    &\geq \mathbb{E}_{p(\bm{f}\mid \bm{\tilde{f}}, \bm{\lambda}, \bm{\theta})} 
    \log \left[p(\{\bm{x}^{(r)}_{1:N}\}\mid \bm{f}, \bm{\lambda}, \bm{\theta}) \right]\\
    &\triangleq \log \tilde{p}(\{\bm{x}^{(r)}_{1:N}\}
    \mid \bm{\tilde{f}}, \bm{\lambda}, \bm{\theta})
    \label{eq:vaele:p_tilde},
\end{align}
where we have reintroduced the superscript notation and used 
$\{\bm{x}^{(r)}_{1:N}\}$ as a shortcut for $\{\bm{x}^{(r)}_{1:N}\}_{r=1}^R$. Manipulating Equation~\eqref{eq:vaele:p_tilde} results in
\begin{align}
    \log \tilde{p}(\{\bm{x}^{(r)}_{1:N}\}\mid \bm{\tilde{f}}, \bm{\lambda}, \bm{\theta}) = 
    \sum_{r=1}^R \Big(&
    \log p(\bm{x}^{(r)}_1) + \sum_{t=1}^{N-1}\sum_{i=1}^d \log \mathcal{N}\left(
        x^{(r)}_{i,t+1}\mid x^{(r)}_{i, t} + \bm{A}_i(\bm{x}^{(r)}_{t})\bm{\tilde{f}}_i, \lambda_i^{-1}\right)\\
        &-\frac{1}{2}\sum_{t=1}^{N-1}\sum_{i=1}^d \bm{P}_i(\bm{x}^{(r)}_{t})\Big),
        \label{eq:vaele:p_tilde_2}
\end{align}
where, just like in Equation~\eqref{eq:vaele:transformation_matrix}, 
$\bm{P}_i(\bm{x}^{(r)}_{t})$ should be read as a scalar. A key property of 
Equation~\eqref{eq:vaele:p_tilde_2} is that it is written as a sum of R terms, each 
related to a different example $\bm{x}^{(r)}_{1:N}$. 
Following \cite{hensman2013gaussian}, we can use this lower bound in the \gls*{sgp} version of the
original generative model as if it was
$\log p(\{\bm{x}^{(r)}_{1:N}\}_{r=1}^R\mid \bm{\tilde{f}}, \bm{\lambda}, \bm{\theta})$, resulting 
in the graphical model shown in Figure~\ref{fig:vaele:sparse_generative}. Note 
that the inducing variables work now as the global variables required by 
\gls*{svi}. Therefore, we shall consider the graphical model 
from Figure~\ref{fig:vaele:sparse_generative} as our complete generative 
model.

\subsection{Variational approximation to the \gls*{sde}-based model
\label{sec:vaele:variational_approx}}
As usual with mean field \glsdesc{vi}, we start by breaking the posterior 
dependencies of the latent variables to achieve a tractable distribution: 
\begin{equation}
    p(\bm{x}_{1:N}, \bm{\tilde{f}}, \bm{\lambda}, \bm{\theta}, D\mid \bm{y}_{1:N}) 
    \approx q(\bm{x}_{1:N}, \bm{\tilde{f}}, \bm{\lambda}, \bm{\theta}, D) =
    q(\bm{x}_{1:N}) q(\bm{\tilde{f}}, \bm{\lambda}, \bm{\theta})q(D).
    \label{eq:vaele:tmp_variational_approx}
\end{equation}

We consider a simple mean field approximation that just factorizes the distributions 
distinguishing between the terms directly modeling the drift and diffusion, 
$\bm{\tilde{f}}$ and $\bm{\lambda}$, and other global variables $\bm{\theta}$:
\begin{equation}
  q(\bm{\tilde{f}}, \bm{\lambda}, \bm{\theta}) = 
  q(\bm{\tilde{f}}, \bm{\lambda}\mid \bm{\theta}) q(\bm{\theta})
  \approx p(\bm{\tilde{f}}, \bm{\lambda}, \bm{\theta} \mid \bm{y}_{1:N}).
  \label{eq:vaele:inducing_points_distro_1} 
\end{equation} 
Note that the variational distributions of $\bm{\tilde{f}}$ and $\bm{\lambda}$ may be 
affected by $\bm{\theta}$. Think for example, in the length-scale parameters of 
the Gaussian kernels. To simplify the variational approximation,
we take the variational factor $q(\bm{\theta})$ to be a Dirac distribution 
$q(\bm{\theta})=\delta_{\bm{\theta}^*}(\bm{\theta})$. 
Equation~\eqref{eq:vaele:inducing_points_distro_1} becomes 
\begin{equation}
  q(\bm{\tilde{f}}, \bm{\lambda}, \bm{\theta}) =
  q(\bm{\tilde{f}}, \bm{\lambda}\mid \bm{\theta} ^ * ) \delta_{\bm{\theta} ^ *}
  (\bm{\theta}).
  \label{eq:vaele:inducing_points_distro_2}
\end{equation}
Since $p(\bm{f}, \bm{\lambda}\mid \bm{\theta}^*)$ was chosen as a conjugate prior in 
Equation~\eqref{eq:vaele:full_model}, the optimal 
$q(\bm{\tilde{f}}, \bm{\lambda}\mid \bm{\theta}^*)$
is also in the same exponential family and therefore, it is the product of $d$
multivariate Gaussian-Gamma distributions:
\begin{equation}
\begin{split}
    q(\bm{\tilde{f}}, \bm{\lambda}\mid \bm{\theta}^*) =&
    \prod_{i=1}^{d} \mathcal{NG}(\bm{\tilde{f}}_i, \lambda_i \mid 
    \bm{\mu}_i, \bm{\Sigma}_i, \alpha_i, \beta_i), \qquad \text{where}\\
    \mathcal{NG}(\bm{\tilde{f}}_i, \lambda_i \mid 
    \bm{\mu}_i, \bm{\Sigma}_i, \alpha_i, \beta_i) \triangleq&
    \mathcal{N}(\bm{\tilde{f}}_i\mid \bm{\mu}_i,\lambda_i^{-1}\bm{\Sigma}_i)\cdot 
    \Gamma(\lambda_i \mid \text{shape}=\alpha_i, \text{rate}=\beta_i).
\end{split}
\end{equation}

Therefore, to find the variational approximation 
$q(\bm{\tilde{f}}, \bm{\lambda}\mid \bm{\theta}^*)$ we only need to determine 
which are the optimal parameters of the Gaussian-Gamma distributions. 
For the moment, let us assume that each $\bm{\tilde{f}}_i$ has well-defined variational parameters 
$[\bm{\mu}_i, \bm{\Sigma}_i, \alpha_i, \beta_i]^T$, although in 
Section~\ref{sec:vaele:lower_learning} we shall introduce a different parametrization.

After studying the variational approach to the \gls*{sde} model, we can plug in 
Equation~\eqref{eq:vaele:inducing_points_distro_2} into 
Equation~\eqref{eq:vaele:tmp_variational_approx} to find the complete 
variational approximation, which is
\begin{equation*}
  \begin{split}
     p(\bm{x}_{1:N}, \bm{\tilde{f}}, \bm{\lambda}, \bm{\theta}, D\mid \bm{y}_{1:N}) 
     & \approx q(\bm{x}_{1:N}, \bm{\tilde{f}}, \bm{\lambda}, \bm{\theta}, D) \\
     & = q(\bm{x}_{1:N}) q(\bm{\tilde{f}}, \bm{\lambda},\bm{\theta})q(D)\\
     & = q(\bm{x}_{1:N}) 
     q(\bm{\tilde{f}}, \bm{\lambda}\mid \bm{\theta} ^ * ) \delta_{\bm{\theta} ^ *}(\bm{\theta})
     q(D).
 \end{split}
\end{equation*}
We can further simplify the variational approximation by assuming that the 
decoding parameters are singular $q(D) = \delta_{D^*}(D)$ and therefore
\begin{equation}
     p(\bm{x}_{1:N}, \bm{\tilde{f}}, \bm{\lambda}, \bm{\theta}, D\mid \bm{y}_{1:N}) 
     \approx q(\bm{x}_{1:N}) 
     q(\bm{\tilde{f}}, \bm{\lambda}\mid \bm{\theta} ^ * ) \delta_{\bm{\theta} ^ *}(\bm{\theta})
     \delta_{D^*}(D).
    \label{eq:vaele:variational_approx}
\end{equation}

A key insight of the \gls*{svae} is that 
Equation~\eqref{eq:vaele:variational_approx} does not break the dependencies 
of $\bm{x}_{1:N}$ across time, which is intended to permit an accurate and 
flexible representation of the dynamics of state space~\cite{johnson2016composing}.
In our case, since the state space is modeled through a \gls*{sde}, the optimal 
factor graph $\bm{x}_{1:N}$ is a Markov chain. In order to help capturing the relevant
dynamical information from $\bm{y}_{1:N}$ an encoding network is used. This permits 
obtaining probabilistic guesses of the latent vector $\bm{x}_t$ from the 
observation $\bm{y}_t$, introducing back-constraints that help in learning the dynamics.

To incorporate both the Markovian dynamics and the encoding information into
the variational distribution $q(\bm{x}_{1:N})$, \cite{johnson2016composing} 
proposes to parametrize it as a conditional random field. 
In our case, the conditional Markov field can be written as 
\begin{align}
  q(\bm{x}_{1:N}) \triangleq q(\bm{x}_{1:N}, E) &\propto\psi(\bm{x}_1) \prod_{t=1}^{N-1}\psi(\bm{x}_t, \bm{x}_{t+1})
        \prod_{t=1}^{N} \psi(\bm{x}_t, \bm{y}_t, E),
    \label{eq:vaele:crf}
\end{align}
where the encoding potentials $\psi(\bm{x}_t, \bm{y}_t, E)$
are Gaussian factors whose means and covariances depend on $\bm{y}_t$ through
an encoder neural network (note the mnemonic $E$). However, it is unrealistic to assume that 
this encoder will be able to output a good latent vector from a single measurement $\bm{y}_t$. What 
$\bm{y}_t$ lacks is the ``context'' of the time series (i.e., its previous 
and future samples), without which it would be impossible to extract the relevant dynamics.

A simple approach to tackle this issue that is connected with the dynamical systems theory would be as follows. 
Since we assume that $\bm{y}_{1:N}$ has been generated by 
some high-dimensional stationary dynamical
system, we could build a delay embedding:
\begin{equation}
    \bm{Y}_t = \Big[\bm{y}_t^T, \bm{y}_{t-\tau}^T, \dots, 
    \bm{y}_{t - (l -1) \tau}^T\Big]^T 
    \qquad \text{for } t=1 + (l-1)\tau, 2 + (l-1)\tau,\dots, N,
    \label{eq:vaele:y_embedding}
\end{equation}
where the vectors $\bm{y}_t$, $\bm{y}_{t-\tau}$, ... are concatenated 
together in a single vector, and where $\tau$ and $l$ should be selected
in such a way that the embedding contains all the information required for 
determining the future of the time series. Therefore, we could
feed the encoder with $\bm{Y}_t$ (instead of $\bm{y}_t$) to calculate the
phase state at time $t$. Furthermore, it is possible to automatize the construction
of the delay embedding using a \gls*{cnn} or a \gls*{rnn}, as proposed in~\cite{eleftheriadis2017identification}:
$$\bm{Y}_{1:N} = \text{CNN}(\bm{y}_{1:N}, E) \qquad\text{or}\qquad\bm{Y}_{1:N} = \text{RNN}(\bm{y}_{1:N}, E).$$
Finally, it is also possible to combine all the methods together, e.g.\, feeding the 
delay embedding into a \gls*{cnn} which then feeds a \gls*{rnn}. We experimentally found 
that feeding the delay embedding into the neural networks may be a valuable way of 
quickly initializing the state space.

Regardless of the specific encoding method, the resulting variational distribution for 
$q(\bm{x}_{1:N})$ now reads
\begin{align}
  q(\bm{x}_{1:N}) \triangleq q(\bm{x}_{1:N}, E) &\propto\psi(\bm{x}_1) \prod_{t=1}^{N-1}\psi(\bm{x}_t, \bm{x}_{t+1})
        \prod_{t=1}^{N} \psi(\bm{x}_t, \bm{Y}_t, E).
    \label{eq:vaele:crf2}
\end{align}
\subsection{Lower bound for SVI and learning phase\label{sec:vaele:lower_learning}}
\paragraph{Lower bound} After selecting an approximation to the true posterior from some tractable 
family, \glsdesc{vi} tries to make this approximation as good as possible by 
reducing the inference procedure to an optimization problem in which the lower 
bound $\mathcal{L}$ of the marginal log-likelihood $\log p(\{\bm{y}^{(r)}_{1:N}\})$
is maximized. Hence, the 
description of the variational approximation cannot be complete without giving 
the lower bound $\mathcal{L}$. Furthermore, to explicitly write the lower bound 
allows us to highlight the role of 
$\tilde{p}(\bm{x}_{1:N}\mid \bm{\tilde{f}}, \bm{\lambda}, \bm{\theta})$ in 
our approximation (see Figure~\ref{fig:vaele:sparse_generative}). We lower bound 
$\log p(\{\bm{y}^{(r)}_{1:N}\})$ as
\begin{equation}
    \begin{split}
      \log p(\{\bm{y}^{(r)}_{1:N}\}) &\geq \mathbb{E}_{
        q(\{\bm{x}^{(r)}_{1:N}\}, E) q(\bm{\tilde{f}}, \bm{\lambda}\mid \bm{\theta}^*)
    }
    \left[ \log \Bigg( 
     \frac{p(\{\bm{y}^{(r)}_{1:N}\}\mid \{\bm{x}^{(r)}_{1:N}\}, D^*)
         p(\{\bm{x}^{(r)}_{1:N}\}\mid \bm{\tilde{f}}, \bm{\lambda}, \bm{\theta}^ *)
         p(\bm{\tilde{f}}, \bm{\lambda}\mid \bm{\theta}^ *)
     }{
     q(\{\bm{x}^{(r)}_{1:N}\}, E)
        q(\bm{\tilde{f}}, \bm{\lambda}\mid \bm{\theta} ^ * )
     }\Bigg)
    \right]\\
    &\geq \mathbb{E}_{
        q(\{\bm{x}^{(r)}_{1:N}\}, E) q(\bm{\tilde{f}}, \bm{\lambda}\mid \bm{\theta}^*)
    }
    \left[ \log \Bigg( 
     \frac{p(\{\bm{y}^{(r)}_{1:N}\}\mid \{\bm{x}^{(r)}_{1:N}\}, D^*)
       \tilde{p}(\{\bm{x}^{(r)}_{1:N}\}\mid \bm{\tilde{f}}, \bm{\lambda}, \bm{\theta}^ *)
         p(\bm{\tilde{f}}, \bm{\lambda}\mid \bm{\theta}^ *)
     }{
     q(\{\bm{x}^{(r)}_{1:N}\}, E)
        q(\bm{\tilde{f}}, \bm{\lambda}\mid \bm{\theta} ^ * )
     }\Bigg)
    \right]\\
    & \triangleq \mathcal{L},
   \label{eq:vaele:lower_bound_1}
\end{split}
\end{equation}  
where we have substituted $p(\bm{x}_{1:N}\mid \bm{\tilde{f}}, \bm{\lambda}, \bm{\theta})$ 
by $\tilde{p}(\bm{x}_{1:N}\mid \bm{\tilde{f}}, \bm{\lambda}, \bm{\theta})$. Equation~\eqref{eq:vaele:lower_bound_1} can be arranged as
\begin{align}
    \mathcal{L}     
    =&\mathbb{E}_{q(\{\bm{x}^{(r)}_{1:N}\}, E)} \left[ \log p(\{\bm{y}^{(r)}_{1:N}\}\mid \{\bm{x}^{(r)}_{1:N}\}, D^*)\right] +
    \mathbb{E}_{q(\{\bm{x}^{(r)}_{1:N}\}, E)}\left[ 
      \mathbb{E}_{q(\bm{\tilde{f}}, \bm{\lambda}\mid \bm{\theta}^*)} \left[
        \log \tilde{p}(\{\bm{x}^{(r)}_{1:N}\}\mid \bm{\tilde{f}}, \bm{\lambda}, \bm{\theta}^ *)
    \right]    \right]\\
     &+ \mathbb{H}\big(q(\{\bm{x}^{(r)}_{1:N}\}, E)\big)
    -\mathcal{KL}\bigg(
      q(\bm{\tilde{f}}, \bm{\lambda}\mid \bm{\theta} ^ *)
      \mid p(\bm{\tilde{f}}, \bm{\lambda}\mid \bm{\theta} ^ *)
    \bigg),
    \label{eq:vaele:lower_bound_2}
\end{align}
where the second term can be written as
\begin{align}
    \mathbb{E}_{q(\bm{\tilde{f}}, \bm{\lambda}\mid \bm{\theta}^*)} \left[ 
      \log \tilde{p}(\{\bm{x}^{(r)}_{1:N}\}\mid \bm{\tilde{f}}, \bm{\lambda}, \bm{\theta}^ *) \right]= \sum_{r=1}^R \Bigg[&
        \log p(\bm{x}^{(r)}_1) + 
         \sum_{t=1}^{N-1}\sum_{i=1}^d \log \mathcal{N}\left(
             x^{(r)}_{i,t+1}\mid x^{(r)}_{i, t} + \bm{A}_i(\bm{x}^{(r)}_{t})\bm{\bm{\mu}}_i, 
         \beta_i/\alpha_i\right)\\
         &-\frac{1}{2}\sum_{t=1}^{N-1}\sum_{i=1}^d \left[\bm{P}_i(\bm{x}^{(r)}_{t})+
     \text{tr}\left(\bm{A}_i(\bm{x}^{(r)}_t)\bm{\Sigma}_i\bm{A}^T_i(\bm{x}^{(r)}_t)\right) \right] \\
         &+ \frac{N-1}{2}\sum_{i=1}^d \left[\psi^0(\alpha_i) - \log (\alpha_i)\right]\Bigg],
    \label{eq:vaele:second_term}
\end{align}
and where $\psi^0(\cdot)$ is the digamma function. Again, we highlight that a key property of the 
lower bound $\mathcal{L}$ is that it breaks as a sum of R terms, each corresponding to a
single $\bm{y}^{(r)}_{1:N}$, which permits the use of mini-batches.

For a better understanding of how the lower bound leads to a proper reconstruction
of the state space, we briefly discuss each of the terms of $\mathcal{L}$ from 
Equations~\eqref{eq:vaele:lower_bound_2} and~\eqref{eq:vaele:second_term}. The first 
term from Equation~\eqref{eq:vaele:lower_bound_2} is basically measuring the expected 
reconstruction error (using a squared error) when decoding the phase space to predict 
$\bm{y}_{1:N}$. Therefore, the first term from Equation~\eqref{eq:vaele:lower_bound_2} 
is encouraging the decoder to fit as close as possible $\bm{y}_{1:N}$, given a variational
distribution of the phase space, $\bm{x}_{1:N}$. This term may also influence the phase space
itself. For instance, it may induce the separation of two points that lie
close in the phase space, say $\bm{x}_{t_1}$ and $\bm{x}_{t_2}$, if they are decoded
into two samples $\bm{y}_{t_1}$ and $\bm{y}_{t_2}$ with very different values.

The second term from Equation~\eqref{eq:vaele:second_term} is responsible for 
adjusting $\{\bm{A}_i(\bm{x}_t)\}_{i=1}^d$ and $\{\bm{\mu}_i\}_{i=1}^d$ so that
$x_{i,t+1}$ can be accurately predicted using $\bm{x}_{t}$. Furthermore, it also
fits $\{\beta_i/\alpha_i\}_{i=1}^d$ to the variance of the residuals that result
from these predictions. The third term from Equation~\eqref{eq:vaele:second_term} 
is more difficult to interpret. The terms $\{\bm{P}_i(\bm{x}_t)\}_{i=1}^d$ come
from Equation~\eqref{eq:vaele:transformation_matrix}, and represent the uncertainty
about the predictions of $f_i(\bm{x}_t)$ using the inducing-points $\bm{\tilde{f}}$. Hence,
Equation~\eqref{eq:vaele:second_term} encourages the minimization of these predictive
uncertainties, which can be achieved by either tuning the hyperparameters of the
kernel or changing the phase space (Note that $\{\bm{P}_i(\bm{x}_t)\}_{i=1}^d$
depends on $\bm{x}_t$). The term involving 
$\big\{\text{tr}\big(\bm{A}_i(\bm{x}_t)\bm{\Sigma}_i\bm{A}^T_i(\bm{x}_t)\big)\big\}_{i=1}^d$
cannot be interpreted in isolation, since it would yield 
$\bm{\Sigma}_i \rightarrow \bm{0}$. This does not occur due to the Kullback-Leibler  
divergence from Equation~\eqref{eq:vaele:lower_bound_2}, which tries to keep
each of the $\bm{\Sigma}_i$ as close as possible to its prior value, say $\bm{\Sigma}_i^{(0)}$.
Therefore, the maximization of $\mathcal{L}$ requires a compromise, which (usually) results
in values of $\bm{\Sigma}_i$ comprised between $\bm{0}$ and $\bm{\Sigma}_i^{(0)}$.
The fourth term from Equation~\eqref{eq:vaele:second_term} favours large values
of $\alpha$, since $\psi(\alpha_i) - \log (\alpha_i)$ is strictly increasing 
for $\alpha_i > 0$. This is consistent with the update rule of the Gaussian-Gamma 
posterior for Gaussians \cite{murphy2007conjugate}, where the $\alpha$
parameter is increased by $n/2$ after the observation of $n$ samples. In our setting,
this could be problematic, since we shall repeatedly update the parameters while
learning the distributions. Again, the solution $\alpha_i \rightarrow \infty$ is avoided
by means of the Kullback-Leibler divergence from Equation~\eqref{eq:vaele:lower_bound_2}.

We have highlighted the key influence of the Kullback-Leibler divergence for the
parameters $\{\bm{\Sigma}_i\}_{i=1}^d$ and $\{\alpha_i\}_{i=1}^d$, but it also
has an effect on the parameters $\{\bm{\mu}_i\}_{i=1}^d$ and $\{\beta_i\}_{i=1}^d$,
forcing a compromise between fitting the data and the prior distribution. The 
last term from Equation~\eqref{eq:vaele:lower_bound_2} to be discussed is 
the entropy (third term). It helps preventing the optimization of $\mathcal{L}$
from simply placing high probability density where the prior does (a random walk 
process), since the entropy encourages to place probability everywhere it is possible.
%
%

\paragraph{Learning phase} Stochastic gradients can be used to learn both the variational distributions, the hyperparameters, and the
encoding and decoding parameters. The computation of the gradients requires calculating the derivates of 
the lower bound from Equation~\eqref{eq:vaele:lower_bound_2}. However, the first two expectation terms
from Equation~\eqref{eq:vaele:lower_bound_2} are hard to compute. Consider, for example, 
the expectations of $\bm{A}_i^{(r)}(\bm{x}_t^{(r)})\bm{\mu}_i$ with respect to 
$q(\{\bm{x}^{(r)}_{1:N}\}, E)$ which, depending on the kernel, may be even impossible
to compute in closed form.
To tackle this issue, a Monte Carlo
approximation shall be used. This requires being able to generate samples from 
the state space $\bm{x}_{1:N}^{(r, s)}$ where 
$\bm{x}_{1:N}^{(r, s)} \sim q(\bm{x}^{(r)}_{1:N}, E) \approx p(\bm{x}^{(r)}_{1:N}\mid~ \bm{y}^{(r)}_{1:N})$. 
The superscript $(r, s)$ identifies one of the $S$ samples from the $r$-th observation used
for the Monte Carlo approximation. Note that to be able to differentiate through
the Monte Carlo approximation, the samples should keep the dependencies with 
the parameters $E$ of $q(\{\bm{x}^{(r)}_{1:N}\}, E)$. 
This can be achieved through the reparametrization trick~\cite{kingma2013auto}. Hence, 
we shall use the notation for the samples as $\bm{x}_{1:N}^{(r,s)}(E)$. 
To draw samples from the phase 
space, we should first infer the optimal variational factors $q(\bm{x}_{1:N}, E)$.
The optimal variational 
distribution $q(\cdot)$ in a local mean field approximation can be found 
by taking expectations with respect 
to the other variational factors \cite{murphy2012machine}. In our case, finding the 
variational factor $q(\bm{x}_{1:N}, E)$ requires fixing the \gls*{sde} parameters
$\bm{\tilde{f}}$ and $\bm{\lambda}$, the hyperparameters $\bm{\theta}$ and the
encoding potentials $\{\psi(\bm{x}_t, \bm{y}_t, E)\}_{t=1}^N$ before computing the 
expectation
\begin{equation}
    \log q(\bm{x}_{1:N}, E) \propto 
    \mathbb{E}_{q(\bm{\tilde{f}}, \bm{\lambda}\mid \bm{\theta}^*)} 
    \left[
        \log \tilde{p}(\bm{x}_{1:N}\mid \bm{\tilde{f}}, \bm{\lambda}, \bm{\theta}^*)
    \right],
    \label{eq:vaele:optimal_q_x_3}
\end{equation}
which permits identifying the following conditional Gaussian factors: 
\begin{equation}
\begin{split}
    &\psi(\bm{x}_1) = p(\bm{x}_1),\\
    &\psi(\bm{x}_t, \bm{x}_{t+1}) = 
    \mathcal{N}\big(\bm{x}_{t+1} \mid \bm{x}_t + \hat{\bm{\mu}}(\bm{x_t}),
    \text{diag}(\bm{\beta}/\bm{\alpha})\big)\qquad \text{with}\\
    &\hat{\bm{\mu}}(\bm{x}_t) = [\bm{A}_1(\bm{x}_t)\bm{\mu}_1, 
    \bm{A}_2(\bm{x}_t)\bm{\mu}_2,\dots,
    \bm{A}_d(\bm{x}_t)\bm{\mu}_d]^T \qquad \text{for } t=1, \dots, T-1,
    \label{eq:vaele:conditional_dyn}
\end{split}
\end{equation}
where the $\bm{A}_i$ has been defined in 
Equation~\eqref{eq:vaele:transformation_matrix}, $\bm{\mu}_i$ is the $i$-th mean of
the $i$-th Gaussian-Gamma distribution from 
$q(\bm{\tilde{f}}, \bm{\lambda}\mid \bm{\theta}^*)$, and 
$\bm{\alpha}$ and $\bm{\beta}$ are the result of stacking the 
$\{\alpha_i\}_{i=1}^d$ and $\{\beta_i\}_{i=1}^d$ parameters from this same 
distribution.  According to Equation~\eqref{eq:vaele:conditional_dyn}, we approximate 
the dynamics of the system as if it evolved following 
\begin{equation}
    \begin{split}
    &\bm{x}_{t+1} = \bm{x}_t + \hat{\bm{\mu}}(\bm{x}_t) + \bm{w}_t,\\
        &\bm{w}_t \sim \mathcal{N}\big(\bm{0}, \text{diag}(\bm{\beta} / \bm{\alpha})\big),
    \label{eq:vaele:variational_dynamics}
\end{split}
\end{equation}
where $\bm{w}_t$ plays the role of a random noise. Note that this variational model 
is analogous to any \gls*{ssm}. Therefore,
it is possible to efficiently draw samples from $q(\{\bm{x}_{1:N}\})$ 
by using a forward filtering pass that collects information about
the observations from the past; and then perform sampling in the backward pass,
using at each node the conditional node potentials and combining them 
with the information from the past of the 
node~\cite[Section~17.4.3]{murphy2012machine}. 

However, closed form solutions of the filtering and smoothing problems are only
available for a few special cases. For example, the well-known Kalman filter 
provides the exact distributions for linear Gaussian systems \cite{kalman1960new}.
In \gls*{gp} dynamical systems, where the transition 
dynamics are governed by \gls*{gps}, several approximations based on moment
matching have been proposed~\cite{candela2003propagation, deisenroth2009analytic, deisenroth2012robust}.
Our solution uses the forward and backward passes described by \cite{deisenroth2009analytic, 
deisenroth2012robust} which not only permits  efficiently drawing samples from the chain, but 
also calculating the required statistics at each node. The main drawback with this 
approximation is this method only works when using squared exponential covariance functions
as kernels~\cite{deisenroth2009analytic,deisenroth2012robust}. As an alternative method that 
enables the use of any kernel, we also implemented a particle filter,  as suggested in~\cite{frigola2013}.

After sampling the state space, a Monte Carlo approximation to the lower bound
$\mathcal{L}$ can be computed. The gradients 
of the neural networks and the hyperparameters $\bm{\theta}$ can be calculated 
using automatic differentiation tools; in our implementation, we have used TensorFlow~\cite{tensorflow}
and the GPflow library~\cite{gpflow}. 

Finally, instead of using standard gradients, we exploit the exponential family 
conjugacy structure of the Gaussian-Gamma distributions to efficiently compute natural gradients 
with respect to the variational parameters. Indeed, the natural gradients can be 
computed explicitly for each dimension, which enables a fast update of the \gls*{sde}
parameters. We note with $\bm{\eta}^{(i)}$  
the natural parameters of the $i$-th Gaussian-Gamma, which are
$$\bm{\eta}^{(i)} = \Big[\eta^{(i)}_0, \eta^{(i)}_1, \left(\bm{\eta}^{(i)}_2\right)^T, 
\left(\bm{\eta}^{(i)}_3\right)^T \Big]^T =
\Bigg[\alpha_i - \frac{1}{2},-\beta_i-\frac{\bm{\mu}_i^T \bm{\Sigma}_i ^{-1} \bm{\mu}_i}{2},
    \bm{\mu}_i^T \bm{\Sigma}_i ^{-1} ,
\text{Vec}\Bigg(-\frac{\bm{\Sigma}_i^{-1}}{2}\Bigg)^T
\Bigg]^T,$$
where $\text{Vec}(\cdot)$ vectorizes the matrix $-\frac{\bm{\Sigma}_i^{-1}}{2}$
so that the natural parameters can be expressed as a vector. The natural gradients with respect 
to $\bm{\eta}^{(i)}$ are:
\begin{align}
  \tilde{\nabla}_{\bm{\eta}^{(i)}}\mathcal{L} =&
  \left[
    (N-1)/2,
    -\sum_{r=1}^R\left[\Delta \bm{x}_i^{(r)}\right]^T\Delta \bm{x}_i^{(r)}/2,
  \sum_{r=1}^R \left[\Delta \bm{x}_i^{(r)}\right]^T\bm{A}_i^{(r)} ,
  -\text{Vec}\left(\sum_{r=1}^R \left[\bm{A}^{(r)}_i\right]^T \bm{A}_i^{(r)}\right)^T /2
  \right]^T\\    
  & - \left(
    \bm{\eta}^{(i)} - \bm{\hat{\eta}}^{(i)}
  \right).
  \label{eq:natural_gradients}
\end{align}
where $\bm{\hat{\eta}}$ are the prior natural parameters and we have noted for convenience
$\Delta \bm{x}_i^{(r)} = \Delta \bm{x}^{(r)}_{i, 1:(N-1)}$ and 
$\bm{A}_i^{(r)}= \bm{A}_i\left(\bm{x}^{(r)}_{1:(N-1)}\right)$.
For a derivation of Equation~\eqref{eq:natural_gradients} see 
Appendix~\ref{app:nat_grad}. 

The resulting \gls*{sde}-\gls*{svae} algorithm for computing the gradients is summarized in 
Algorithm~\ref{alg:vaele:svae}~\cite{johnson2016composing}.

\begin{algorithm}[t]
      \caption{
        \gls*{svae} algorithm for \gls*{sde}-based dynamics. To keep the notation uncluttered, 
        we omit the indices that generate the sets, e.g\., we use 
        $\{\psi(\bm{x}^{(r)}_t, \bm{y}^{(r)}_t, E)\}$
        instead of
        $\{\psi(\bm{x}^{(r)}_t, \bm{y}^{(r)}_t, E)\}^{t=N,r=R}_{t=1,r=1}.$
        \label{alg:vaele:svae}
    }
    \begin{algorithmic}[1]
      \INPUT{A batch of experimental time series $\{\bm{y}^{(r)}_{1:N}\}$; 
            the current variational parameters $\bm{\tilde{\eta}}$ and the 
            hyperparameters $\bm{\theta^*}$ characterizing the distribution 
            $q(\bm{\tilde{f}}, \bm{\lambda}, \bm{\theta}^*)$; and the current
            weights of the encoder and the decoder, $E$ and $D^*$.
        }
        \Function{compute\_gradients}{$\{\bm{y}^{(r)}\}$, $\bm{\tilde{\eta}}$, $\bm{\theta}^*$, $D^*$, $E$}
        \LeftComment{Get the encoding potentials using the encoding network}
        \State{ 
          $\{\psi(\bm{x}^{(r)}_t, \bm{y}^{(r)}_t, E)\}\gets$ 
            \Call{NNet}{$\{\bm{y}_{1:N}^{(r)}\},
            E$}
        }
        \State{
            $\{\bm{\mu}_i,\bm{\Sigma}_i, \alpha_i, \beta_i\}$ = 
        \Call{unpack\_Gaussian\_Gamma\_params}{$\bm{\tilde{\eta}}$}
        }
        \LeftComment{Define the dynamic potentials using Equation~\eqref{eq:vaele:conditional_dyn}}
        \ForAll {pairs $(r, t)$}
        \State{
          $\hat{\bm{\mu}}(\bm{x}^{(r)}_t) \gets [\bm{A}_1(\bm{x}^{(r)}_t)\bm{\mu}_1, 
          \bm{A}_2(\bm{x}^{(r)}_t)\bm{\mu}_2,\dots,
          \bm{A}_d(\bm{x}^{(r)}_t)\bm{\mu}_d]^T$
          \State{
            $\psi(\bm{x}^{(r)}_t, \bm{x}^{(r)}_{t+1}) \gets 
            \mathcal{N}\Big(\bm{x}^{(r)}_{t+1} \mid \bm{x}^{(r)}_t + \hat{\bm{\mu}}(\bm{x}^{(r)}_t),
            \text{diag}(\bm{\beta}/\bm{\alpha})\Big)$
          }
        } 
        \EndFor
        \LeftComment{
            Draw $S$ samples from $q(\bm{x^{(r)}}_{1:N}, E)$ and compute its entropy 
            $\mathbb{H}\big(q(\bm{x}^{(r)}_{1:N}, E)\big)$
        }
        \State{$\{\bm{\hat{x}}_{1:N}^{(r,s)}\}$,
          $\{\mathbb{H}\big(q(\bm{x}^{(r)}_{1:N}, E)\} \gets$ \big( \par
                \hskip\algorithmicindent
                \Call{sample\_and\_get\_entropy}{
                  $\{\psi(\bm{x}^{(r)}_t, \bm{x}^{(r)}_{t+1}, E)\},
                 \{\psi(\bm{x}^{(r)}_t, \bm{y}^{(r)}_t, E)\}$
            }
            \par $\big)$
        }
        \LeftComment{Compute lower bound with Equation~\eqref{eq:vaele:lower_bound_1}}
        \State{
          $\mathcal{L}$ = \Call{compute\_L}{
            $
            \{\bm{y}^{(r)}_{1:N}\},
            \{\bm{y}^{(r)}_{1:N}\},
            \{\mathbb{H}\big(q(\bm{x}^{(r)}_{1:N}, E)\big)\},
            \{\bm{\mu}_i,\bm{\Sigma}_i, \alpha_i, \beta_i\}
            $
          }
        }
        \LeftComment{Return gradients and natural gradients with Equation~\eqref{eq:natural_gradients}}
        \State{
            \Return{
                $\nabla_{D^*, E,\bm{\theta}^*}\mathcal{L},\;
                \tilde{\nabla}_{\bm{\eta}}\mathcal{L}$
            }
        }
        \EndFunction
    \end{algorithmic}
\end{algorithm}

\section{Validation\label{sec:vaele:validation}}
In this section, we apply our model to a variety of experiments, to illustrate 
its potential to capture salient features of experimental time series. The 
experiments increase in complexity, and each of them tries to illustrate different
aspects of the algorithm.

The experiments were run sharing a similar experimental setup. The encoder 
consisted of a single layer of 1D convolutions (with kernel size of 5 and 32 feature maps), followed by a
single \gls*{gru} of size 64 and two \gls*{mlp} (one encoding the mean and the other 
one encoding the scale). The \gls*{mlp} used 128 hidden units with leaky-ReLu activations. 
The convolution layer was fed with five-dimensional delay embeddings, in which the time lag was 
selected using the average mutual information criterion. The delay embeddings were normalized 
before feeding them to the \gls*{sde}-\gls*{svae}. We used both the moment matching 
algorithms and the particle filtering for drawing samples from 
the state space, but no significant differences in the final results were found. For illustrative
purposes we show the moment matching results in the experiments from Sections~\ref{sec:vaele:oscillator}
and~\ref{sec:vaele:lotka}, and the particle filter results in the experiments from 
Sections~\ref{sec:vaele:rossler} and~\ref{sec:vaele:lorenz}. The moment matching 
experiments used squared exponential kernels (as required by the algorithm) and the 
particle filter experiments used a Mattern $3/2$
kernel. The neural networks and the hyperparameters were optimized using the Adam optimizer
with learning rate $3\cdot10^{-4},$ whereas that the natural parameters were optimized 
using vanilla gradient descent with learning rate $0.1$.

\subsection{Excited oscillator\label{sec:vaele:oscillator}}
We begin with a simple example in which a deterministic one-dimensional time 
series $\bm{y}_{1:N}$ oscillates while slowly increasing its amplitude. Specifically,
we generated several realizations of the model using  
\begin{equation}
    \begin{split}
        y^{(r)}_t = \sqrt{2} \exp{\left(0.002\cdot t\right)} 
        \cos\left(\frac{2 \pi}{100}t + \phi^{(r)}\right), 
    \qquad &r=1,2, \dots, 120; \qquad t = 1, 2, \dots, 500
    \end{split}
\end{equation}
where $\phi^{(r)}$ is a random sample generated from a uniform distribution 
~$\mathcal{U}(0, 2\pi)$. Note that this is the only source of randomness of the 
system. 
This time series is quite challenging for our model 
since the diffusion term should vanish to fit the deterministic equation 
generating the data ($\beta/\alpha \rightarrow 0$).  However, the smoothness 
of the signal makes it easier to interpret the figures from the \gls*{svae} 
algorithm, making it a good illustrative example.

\begin{figure}
    \centering
    \begin{subfigure}[t]{0.5\textwidth}
        \centering
        \includegraphics[width=\textwidth]{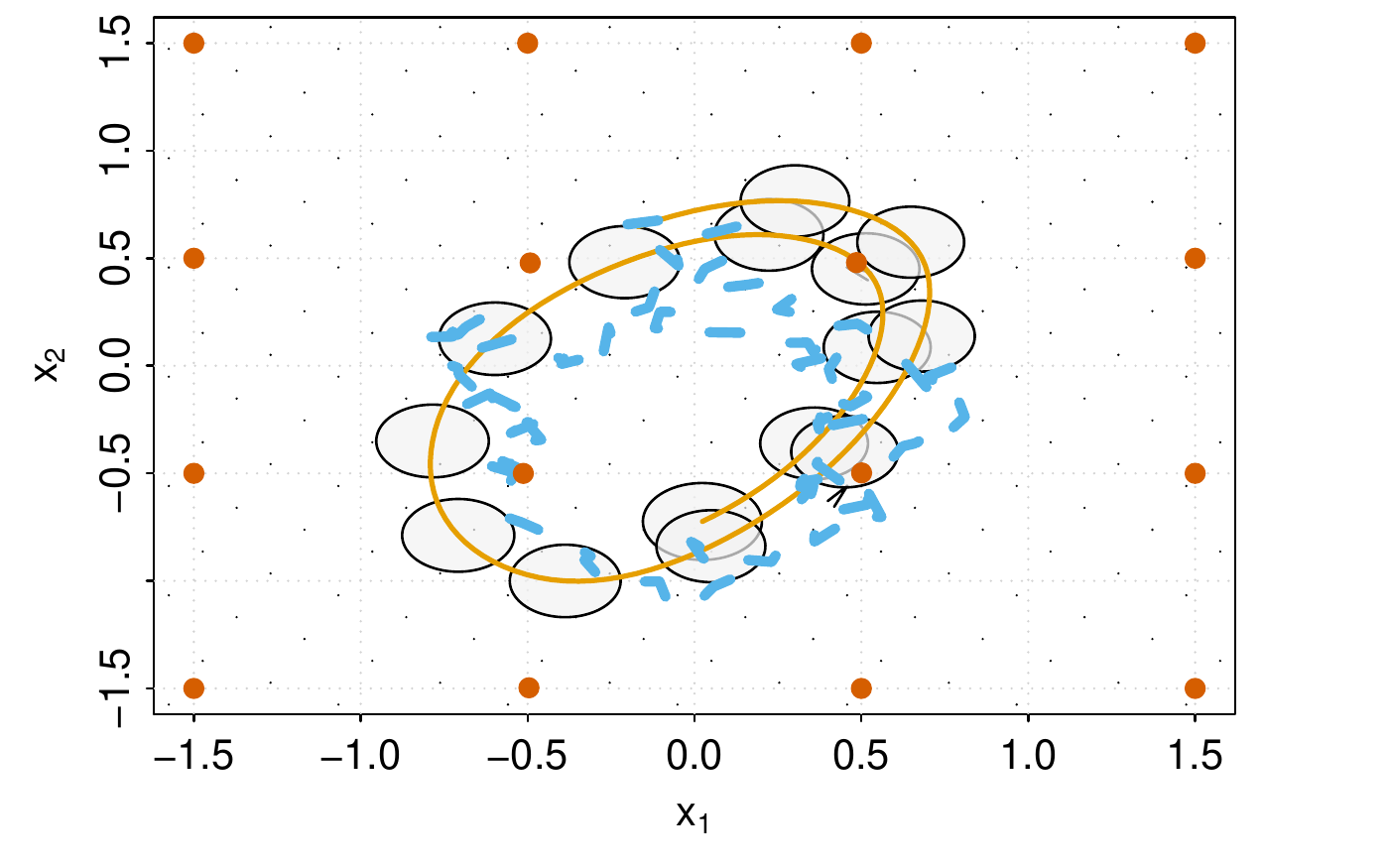}
        \caption{\label{fig:vaele:oscillator_1}}
    \end{subfigure}%
    \begin{subfigure}[t]{0.5\textwidth}
        \centering
        \includegraphics[width=\textwidth]{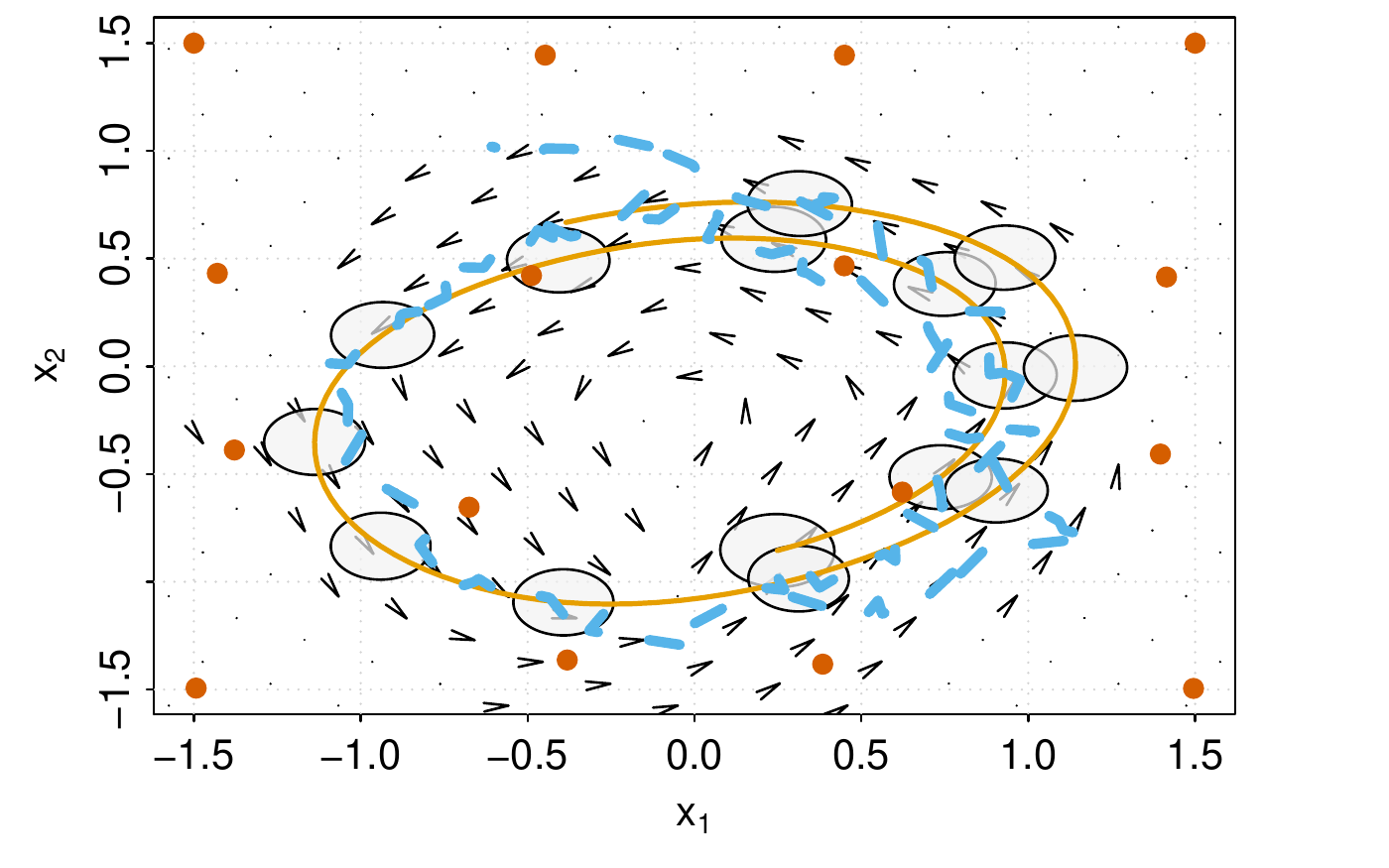}
        \caption{\label{fig:vaele:oscillator_2}}
    \end{subfigure}

    \begin{subfigure}[t]{0.5\textwidth}
        \centering
        \includegraphics[width=\textwidth]{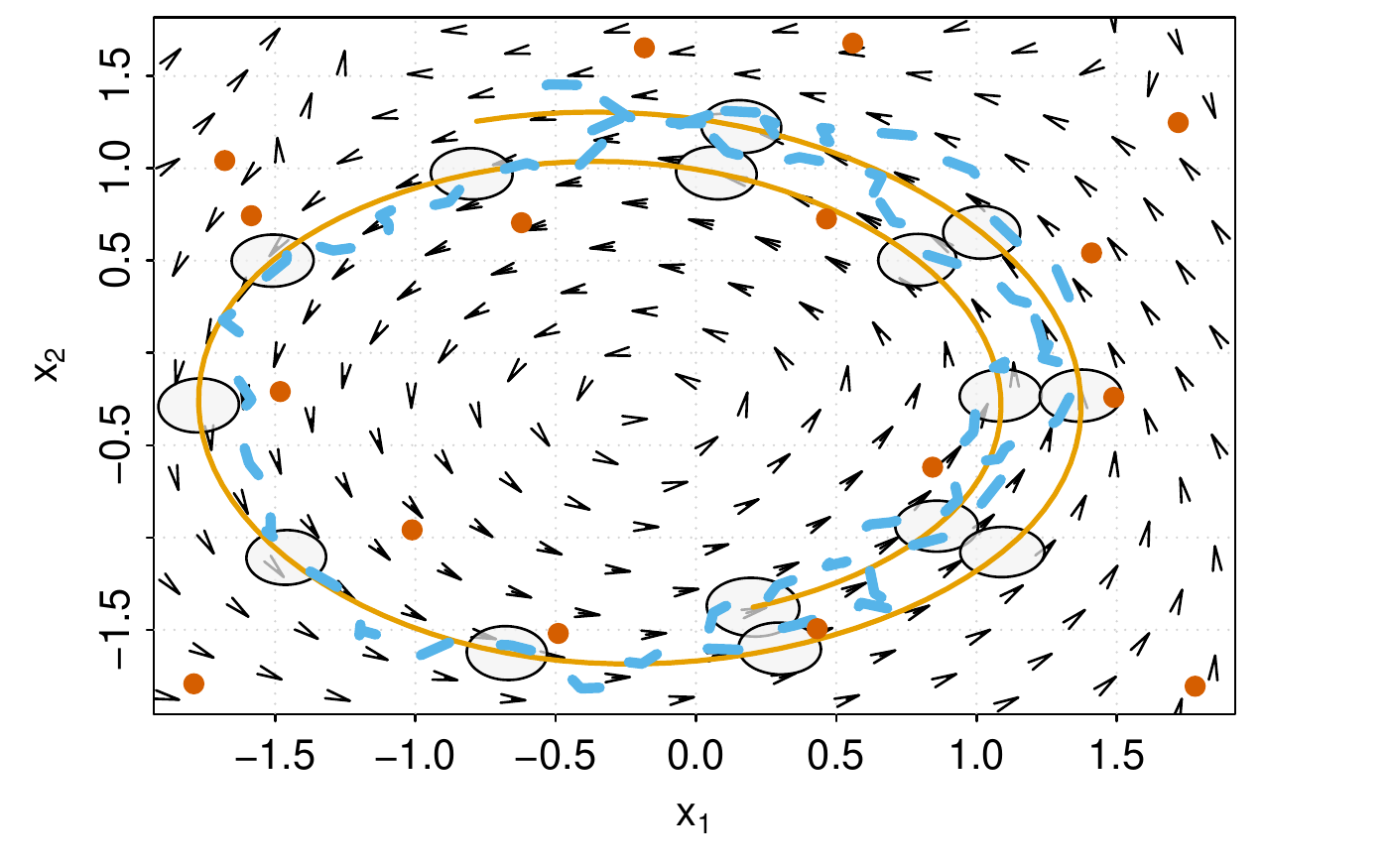}
        \caption{\label{fig:vaele:oscillator_3}}
    \end{subfigure}%
    \begin{subfigure}[t]{0.5\textwidth}
        \centering
        \includegraphics[width=\textwidth]{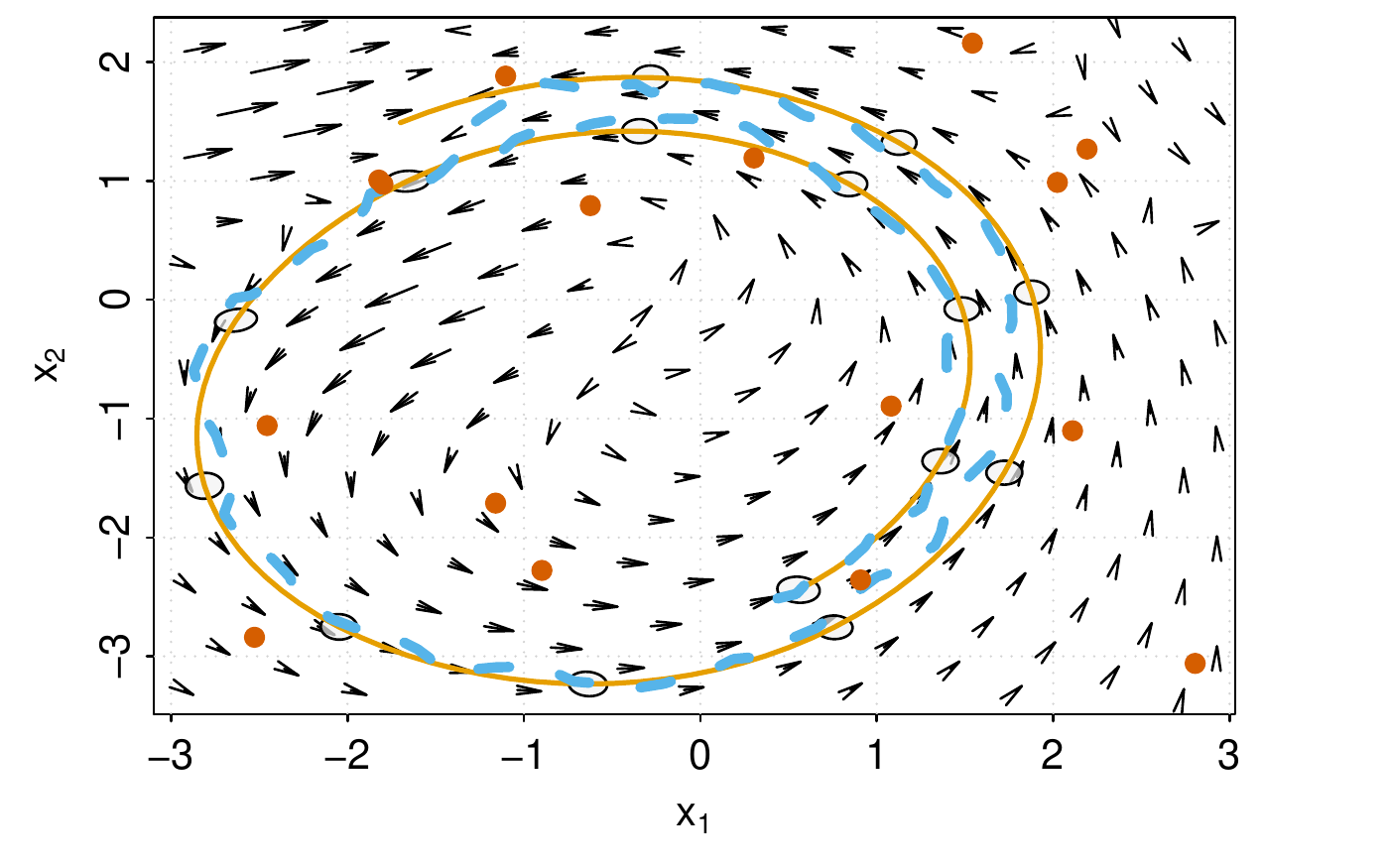}
        \caption{\label{fig:vaele:oscillator_4}}
    \end{subfigure}

    \begin{subfigure}[t]{0.5\textwidth}
        \centering
        \includegraphics[width=\textwidth]{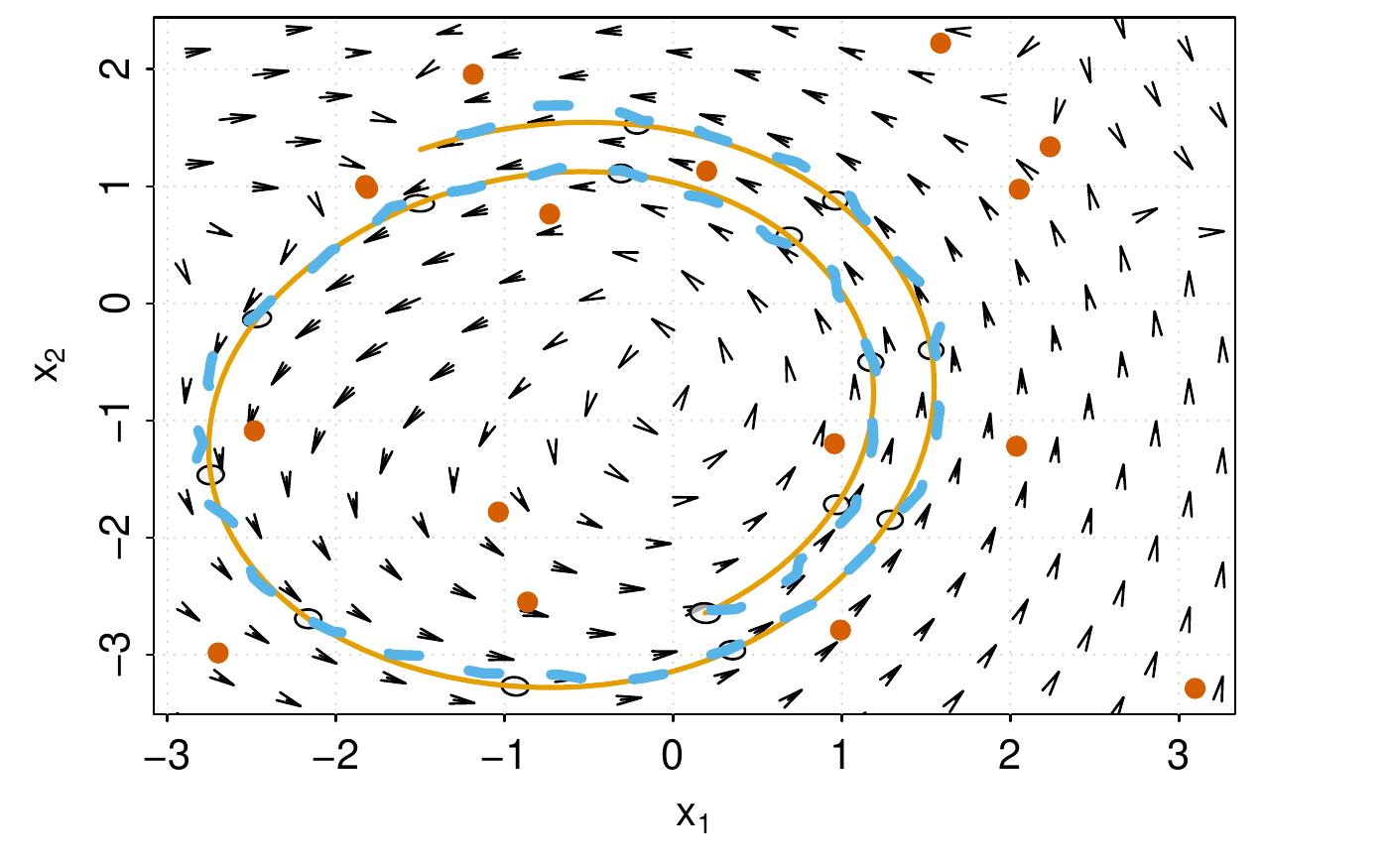}
        \caption{\label{fig:vaele:oscillator_5}}
    \end{subfigure}%
    \begin{subfigure}[t]{0.5\textwidth}
        \centering
        \includegraphics[width=\textwidth]{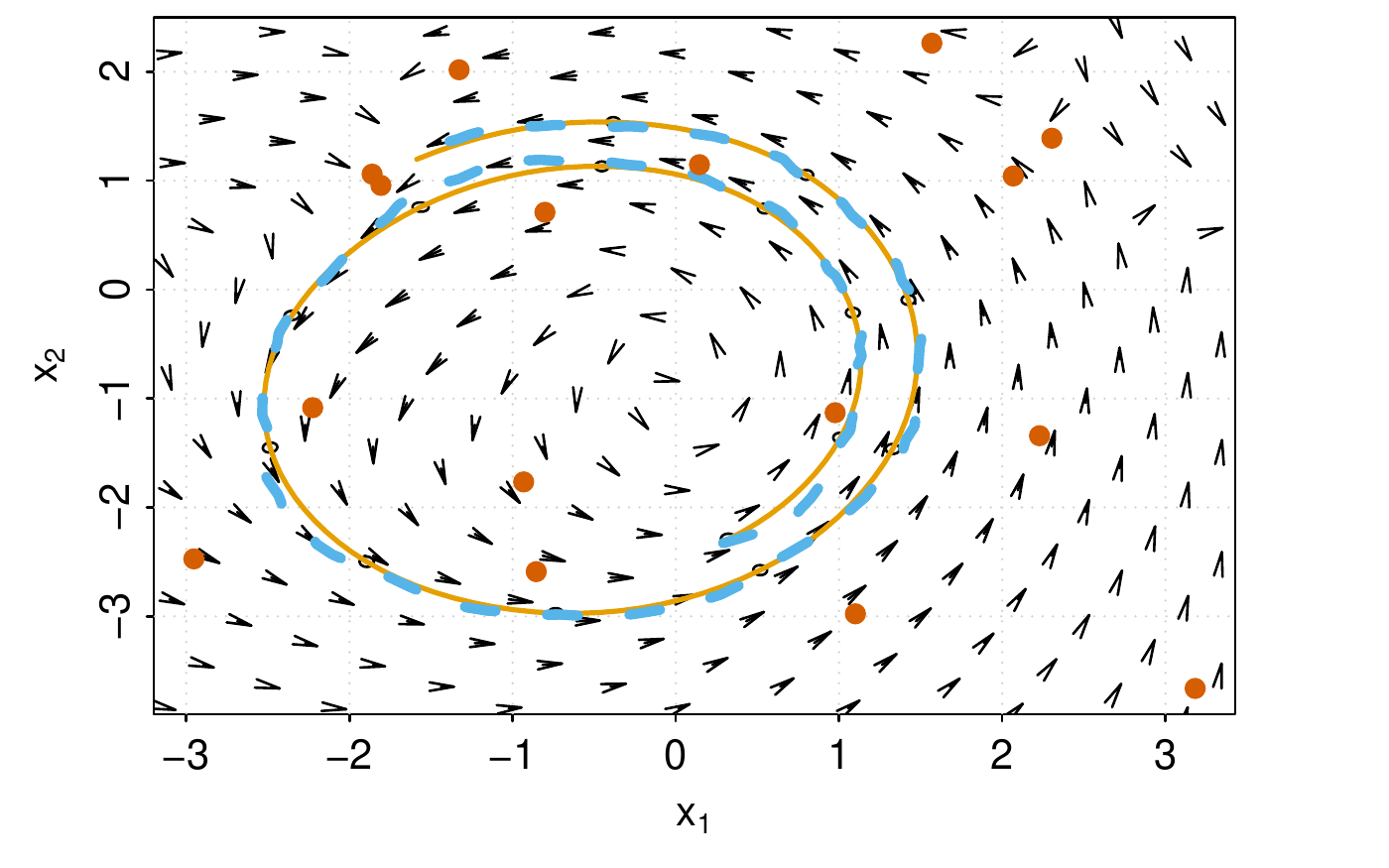}
        \caption{\label{fig:vaele:oscillator_6}}
    \end{subfigure}
    \caption[Evolution of the excited oscillator state space]{Evolution of the state space after 
        (a) 35 iterations, (b) 131 iterations, (c) 323 iterations, (d) 1163 
        iterations, (e) 1847 iterations, and (f) 4247 iterations. The mean 
        and covariances of the encoding network are represented by orange 
        lines and gray ellipses, respectively. A single illustrative sample 
        of the state space is shown with a dashed blue line. Finally, 
        the pseudo-inputs are represented with reddish points, whereas 
        the drift is represented by arrows. 
    \label{fig:vaele:oscillator_ps}}
\end{figure}

Besides the choices for the neural networks described at the beginning of the section,
the following parameters were selected. We used $d=2$ as dimension of the phase 
space; the size of the graph was set to $N=150$, effectively limiting the 
backpropagation through time to $150$ steps; the Monte Carlo approximations were calculated using
$S=10$ samples; and $16$ inducing points were employed, initializing them 
after 5 epochs of training and scattering them into an uniform grid.

Figure \ref{fig:vaele:oscillator_ps} shows a small segment of the state space
at several stages of the training procedure. Note that the state space shows
the representation generated by the 
encoding network and the dynamics fitted by the \gls*{sde}
(Figure~\ref{fig:vaele:oscillator_ps}). A single illustrative sample of 
$\bm{x}_{1:N}$ is also shown. It is worth noting that the state space 
is able to capture the key features of the time series $\bm{y}_{1:N}$: the 
oscillatory pattern and the slow change in amplitude, therefore constructing 
a meaningful state space.

\begin{figure}
  \begin{subfigure}[t]{0.5\textwidth}
    \centering
    \includegraphics[width=\textwidth]{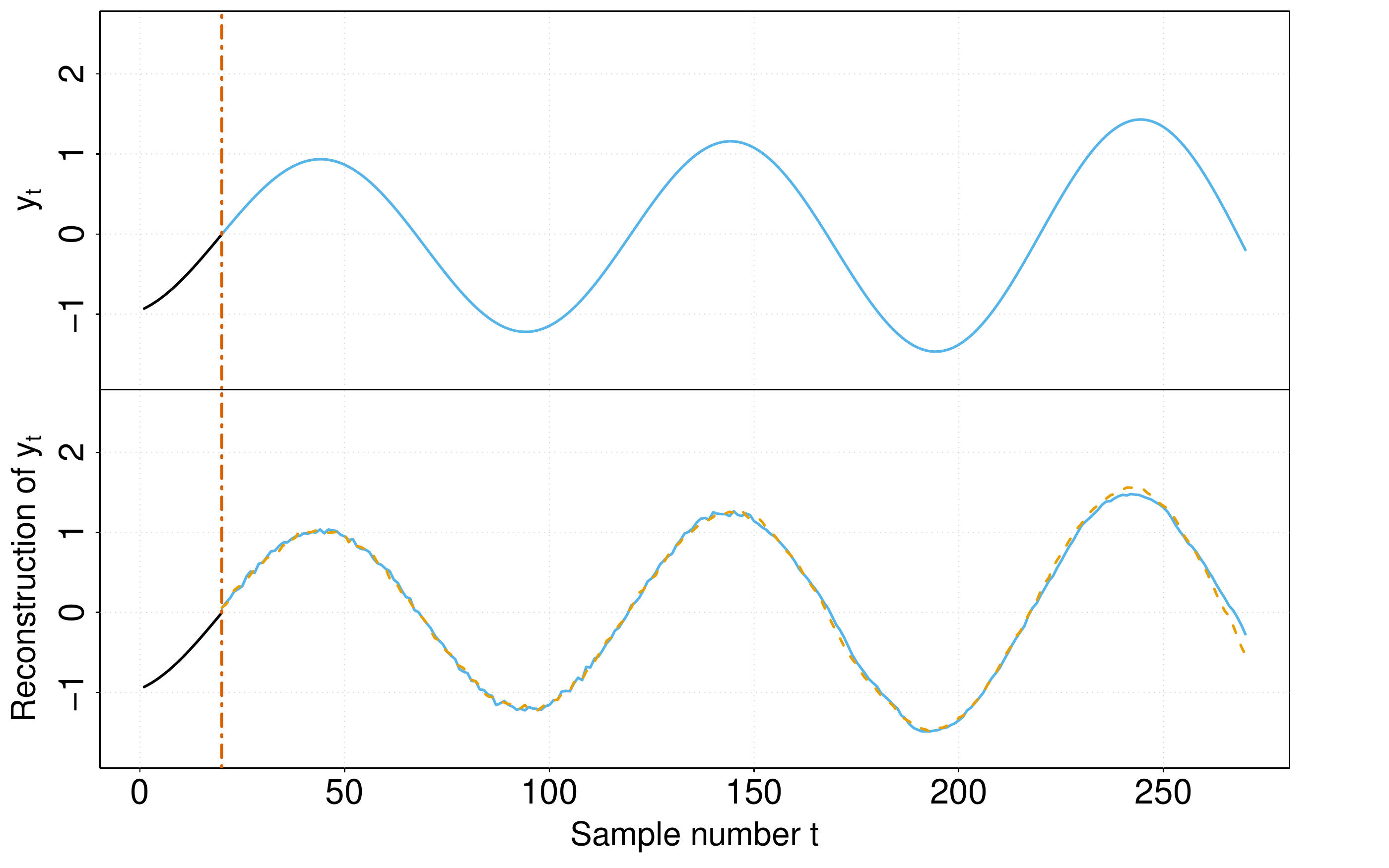}
    \caption{\label{fig:vaele:oscillator_reconstruction}}
  \end{subfigure}
  \begin{subfigure}[t]{0.5\textwidth}
    \centering
    \includegraphics[width=\textwidth]{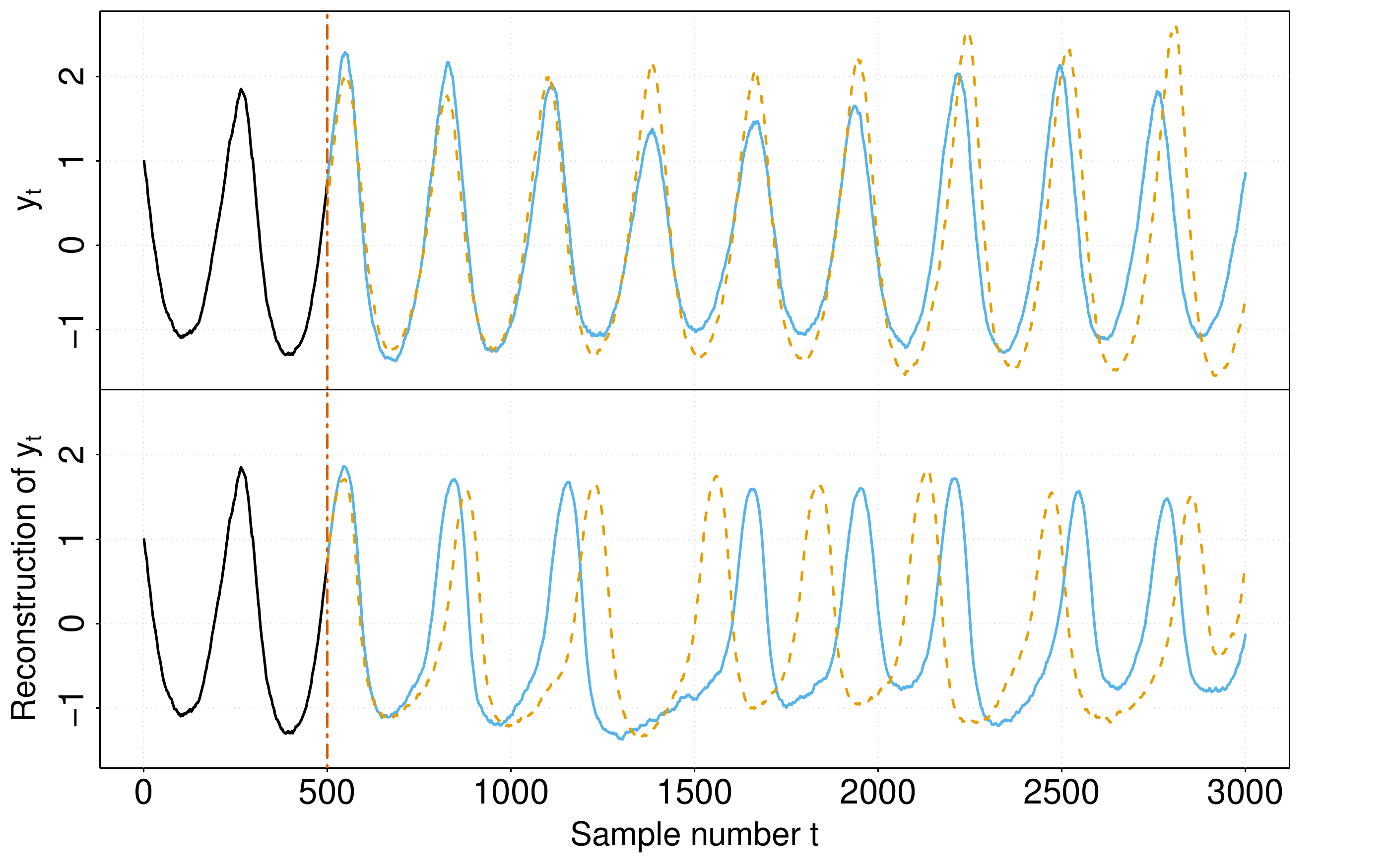}
    \caption{\label{fig:vaele:lokta_volterra_reconstruction}}
  \end{subfigure}
  \caption{
    (a) Predictions from a \gls*{sde}-\gls*{svae} for the excited oscillator. 
    (b) Predictions from a \gls*{sde}-\gls*{svae} for the Lotka-Volterra model. The top
    panels show real data; In (b), two different realizations of the synthetic data are shown,
    given the stochasticity of the model. The bottom panels show synthetic
    time series generated by the \gls*{svae} given the data up to the 
    vertical line, colored in black.
  \label{fig:vaele:reconstructions_1}
}
\end{figure}

To confirm that the dynamics has been properly learned, we generate predictions
from the \gls*{sde}-\gls*{svae} given some small segment of data as input.
This is illustrated in Figure~\ref{fig:vaele:oscillator_reconstruction}.
To generate a new prediction, the encoder creates a trajectory in state space
for the given time series (colored in black, up to the vertical line). When the encoder 
runs out of input data, the last state of the encoded trajectory is used as the
initial state of the fitted \gls*{sde}. The \gls*{sde} generates a new trajectory 
without the intervention of the encoder. Finally, the decoder maps the new phase
space trajectories to the original space, generating new synthetic time series.
Note that, since the \gls*{sde} is a stochastic model, different trajectories may 
result from the same initial state. In Figure~\ref{fig:vaele:oscillator_reconstruction},
two different synthetic time series are shown (in dashed-orange and solid-blue).  
Since the original time series is deterministic, the differences between both
samples are small, although noticeable. This was expected, since the \gls*{sde}-\gls*{svae}
can not reach the limit $\beta/\alpha \rightarrow 0$, and therefore deterministic series
are approximated with \glspl*{sde} with a small diffusion term.

\subsection{Lotka-Volterra\label{sec:vaele:lotka}}
In this section, we apply our \gls*{sde}-\gls*{svae} to a nonlinear 
stochastic model inspired by the Lotka-Volterra equations. These equations
are used in the study of biological ecosystems to
describe the interactions between a predator and its prey. The diffusion 
approximation of the Lotka-Volterra model, described in
\cite{andrieu2010particle, boys2008bayesian}, is given by
\begin{equation}
\begin{split}
    \begin{bmatrix} 
        d x_1(t) \\
        d x_2(t) 
    \end{bmatrix} = &
    \begin{bmatrix} 
        {\alpha}x_1(t) - {\beta}x_1(t)x_2(t) \\
    {\beta}x_1(t)x_2(t) - {\gamma}x_2(t) \end{bmatrix}\, \mathrm{d} t
    \\&+
    \begin{bmatrix} 
        {\alpha}x_1(t) + {\beta}x_1(t)x_2(t) & -{\beta}x_1(t)x_2(t) \\
        -{\beta}x_1(t)x_2(t) & {\beta}x_1(t)x_2(t) + {\gamma}x_2(t)
    \end{bmatrix}^{1/2}
    \begin{bmatrix} 
        d W_1(t) \\
        d W_2(t)
    \end{bmatrix}.
\end{split}
    \label{eq:vaele:lokta_volterra_model}
\end{equation}
Note that the diffusion matrix does not meet the Lamperti transformation criteria 
from Equation~\eqref{eq:lamperti_condition}. In this experiment we test how this 
affects the state space built by the \gls*{svae}.

\begin{figure}
    \centering
    \begin{subfigure}[t]{0.5\textwidth}
        \centering
        \includegraphics[width=\textwidth]{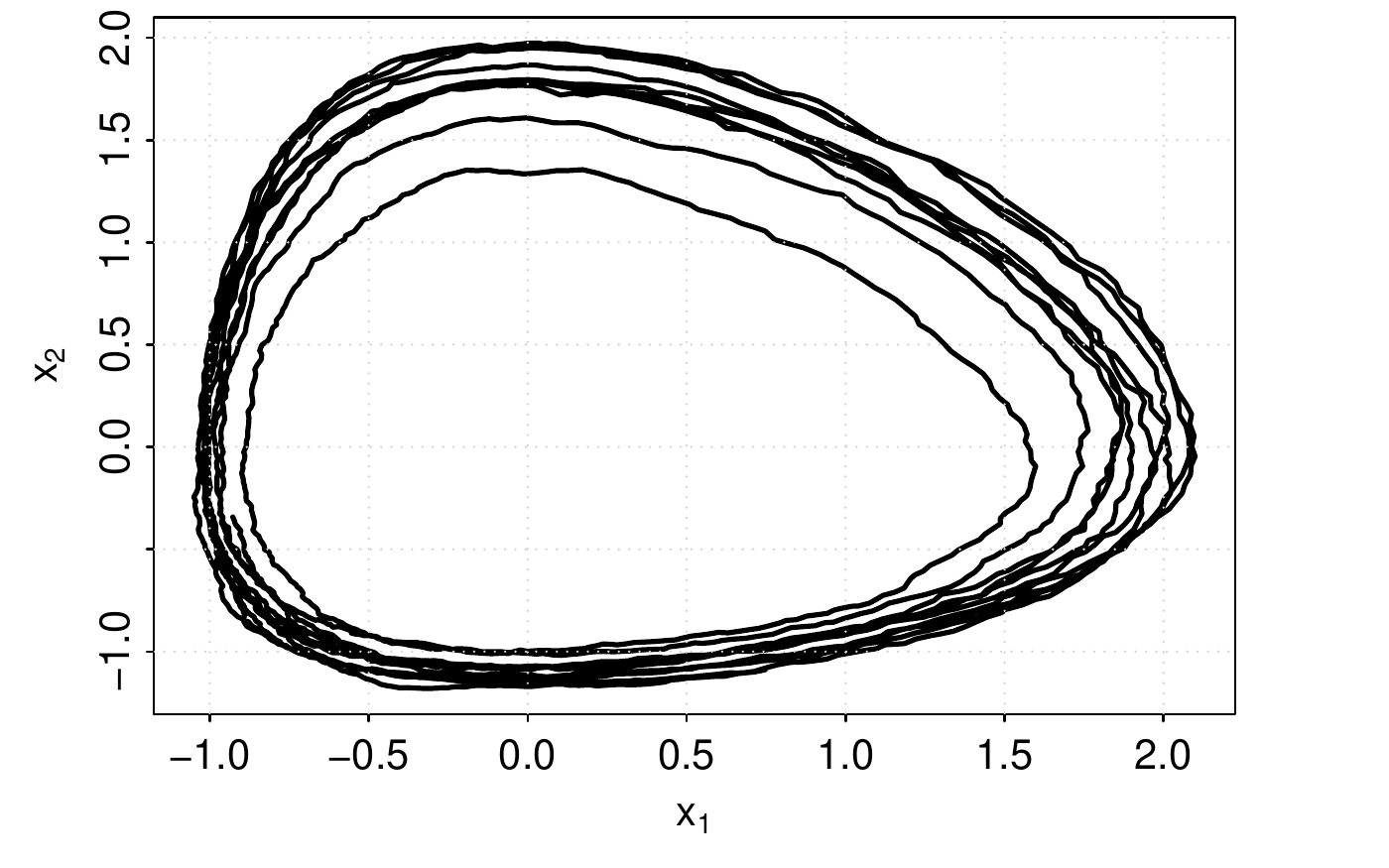}
        \caption{\label{fig:vaele:lokta_volterra_1}}
    \end{subfigure}%
    \begin{subfigure}[t]{0.5\textwidth}
        \centering
        \includegraphics[width=\textwidth]{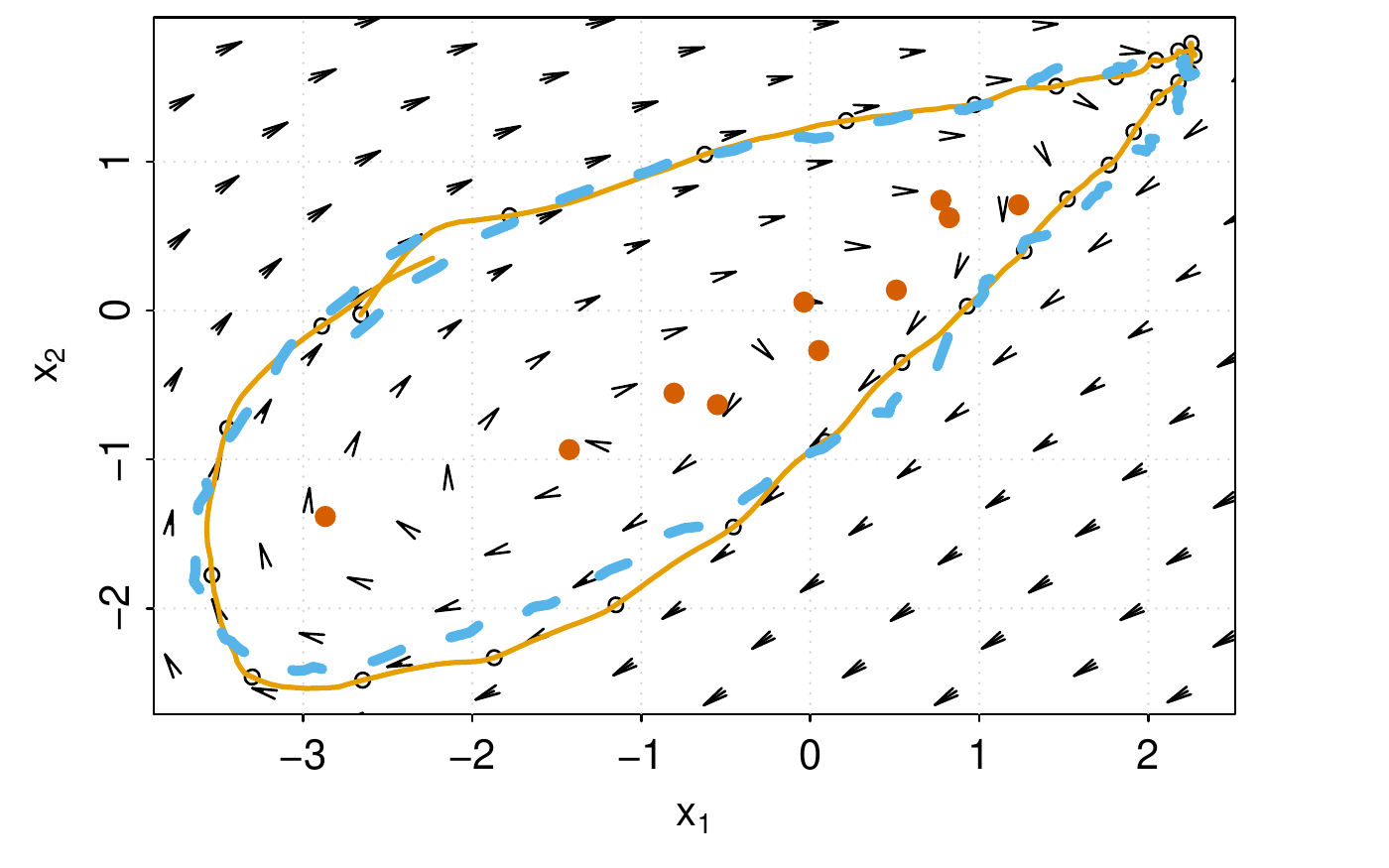}
        \caption{\label{fig:vaele:lokta_volterra_6}}
    \end{subfigure}
    \caption{
      Comparison of the (a) true stochastic Lotka-Volterra state space and (b) the reconstructed 
      one after 450 iterations. The meaning of the symbols from (b) is the same as in 
      Figure~\ref{fig:vaele:oscillator_ps}.
    \label{fig:vaele:lokta_volterra_ps}}
\end{figure}

We analyse the case $\alpha=30$, $\beta=0.2$ and $\gamma=18$. An illustrative
portion of the real state space of the system is shown in 
Figure~\ref{fig:vaele:lokta_volterra_ps}. The \gls*{svae}
was fed with 10 different realizations of the $x_1(t)$ signal, sampled at
1000 Hz. Each of the resulting input signals, 
$y^{(r)}_{t} = x^{(r)}_1(t / 1000)$ had a length of $600$ samples. We used $d=2$ 
for the state space dimension, $N=150$ for the size of the graphical
model, $S=10$ for the number of Monte Carlo samples, and $m=10$ for the number
of inducing points. 

Figure~\ref{fig:vaele:lokta_volterra_ps} compares the true state space of the 
Lotka-Volterra system with the one obtained from the SDE-SVAE at the 
end of the training. Figure~\ref{fig:vaele:lokta_volterra_reconstruction} shows predictions
generated by a trained \gls*{sde}-\gls*{svae}. Both figures illustrate the capability of 
the model to capture the most salient features of the system in state space and to 
generate credible synthetic signals despite the fact that the underlying model does not meet the 
Lamperti transformation criteria. However, the synthetic signals do not fully reproduce all the properties 
of the original system. For example, the variability of the height of the peaks is smaller in the 
synthetic signals, and they do not reach the most extreme values of the original signal.

\subsection{Stochastic Rossler Model\label{sec:vaele:rossler}}
\begin{figure}
    \centering
    \begin{subfigure}[t]{0.5\textwidth}
        \centering
        \includegraphics[width=\textwidth]{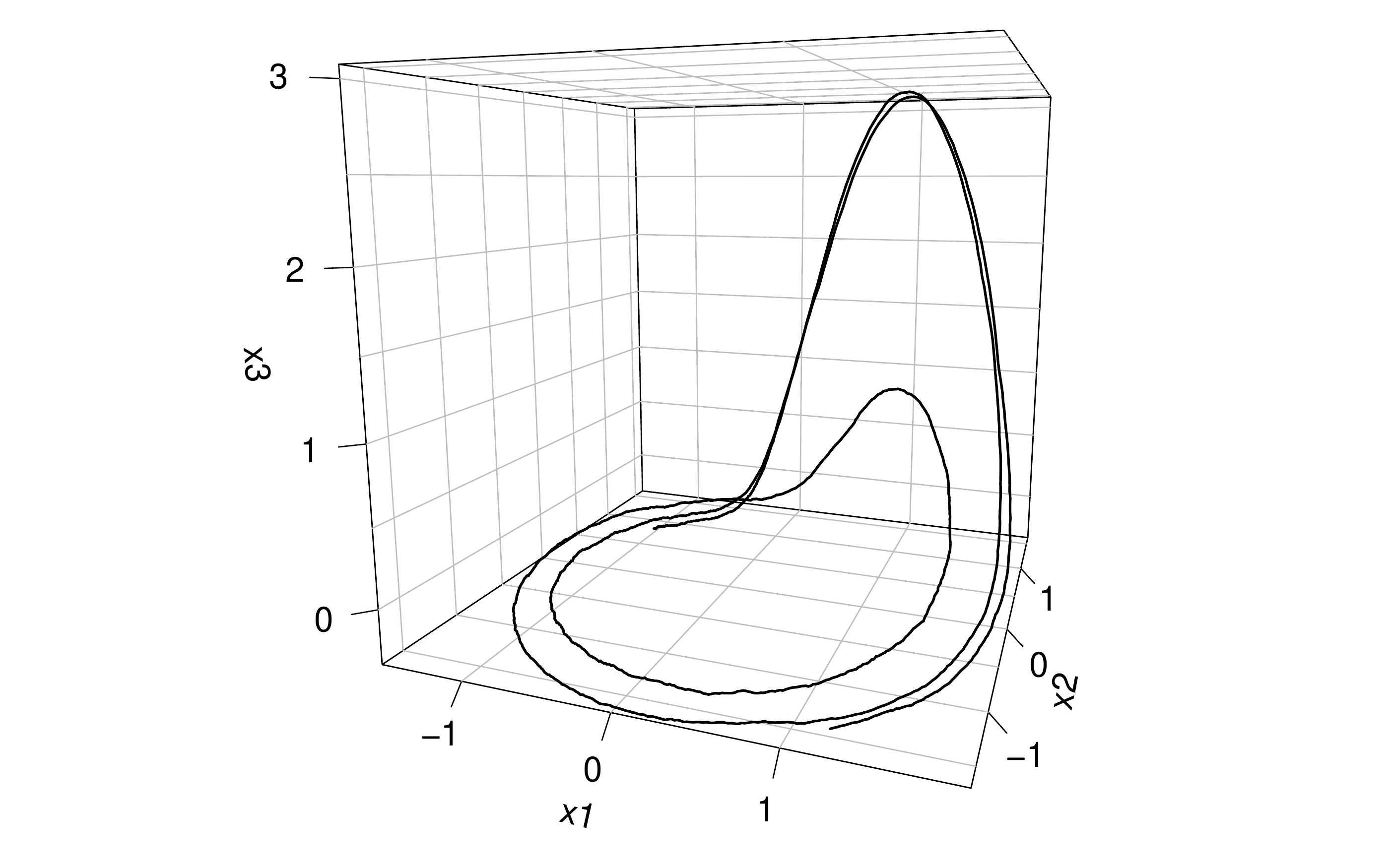}
        \caption{\label{fig:vaele:real_rossler}}
    \end{subfigure}%
    \begin{subfigure}[t]{0.5\textwidth}
        \centering
        \includegraphics[width=\textwidth]{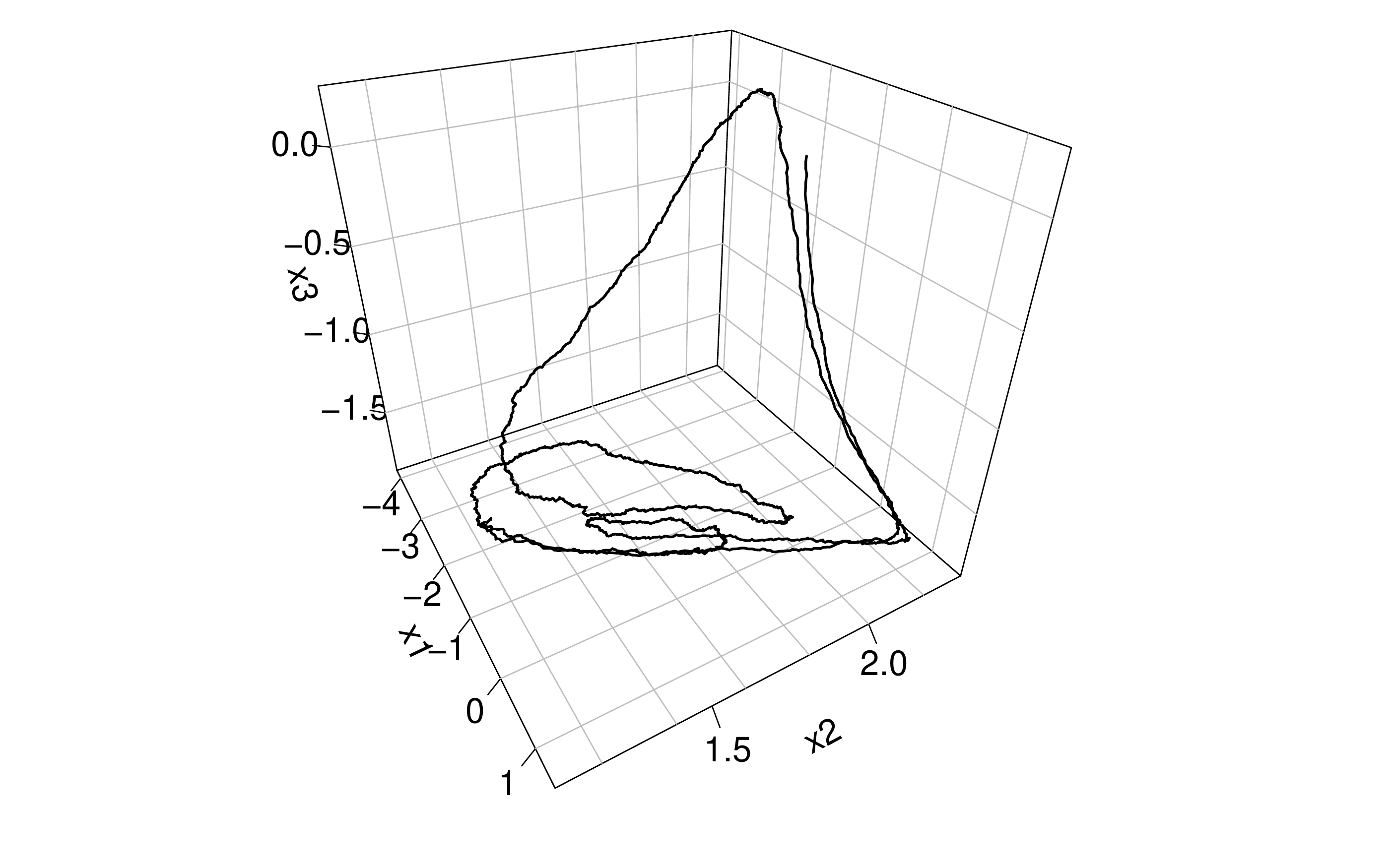}
        \caption{\label{fig:vaele:recons_rossler}}
    \end{subfigure}
    \caption{
      Comparison of the (a) true stochastic Rossler state space and (b) a reconstructed phase space after
      5000 iterations and the observation of experimental data.
    \label{fig:vaele:rossler_ps}}
\end{figure}

In this section, we show a test where the \gls*{sde}-\gls*{svae} did not perform well using a 
stochastic version of the Rossler system:
\begin{equation}
\begin{split}
  dx_1 &= -(x_2 -x_3)dt + \delta \cdot dW_1(t), \\
  dx_2 &= (x_1+ \alpha x_2) + \delta \cdot dW_2(t), \\
  dx_3 &= (\beta + x_3(x_1-\gamma))dt + \delta \cdot dW_3(t),
    \label{eq:vaele:rossler}
\end{split}
\end{equation}
with $\alpha=0.2$, $\beta=0.2$, $\gamma=5.7$ and $\delta=0.1$. A
representative portion of the real state space of the system is shown in 
Figure~\ref{fig:vaele:real_rossler}.  The \gls*{svae} was fed with 200 different 
realizations of the $x_1(t)$ signal, which was sampled at 1000 Hz and chunked in segments
of $100$ samples. We used $d=3$ for the state space dimension, $N=100$ for
the size of the graphical model, $S=10$ for the number of Monte Carlo samples, 
and $m=64$ for the number of inducing points. The inducing points were initialized at the beginning of 
the training using a grid-based approach in the range $[-3, 3]$.

The real and synthetic state space are shown in Figures~\ref{fig:vaele:real_rossler} 
and~\ref{fig:vaele:recons_rossler}, respectively; the predictions generated after 
training the \gls*{sde}-\gls*{svae} are shown in Figure~\ref{fig:vaele:rossler_reconstruction}.
It is possible to visually assess that the state space from Figure~\ref{fig:vaele:real_rossler} 
has a smaller diffusion than the state space from Figure~\ref{fig:vaele:recons_rossler}. As
a consequence, the synthetic reconstructions shown in~Figure~\ref{fig:vaele:rossler_reconstruction}
aren't as smooth as the real time series. Furthermore, the synthetic samples always decay 
to small amplitudes, without showing the rich dynamics of the original system. Clearly, the 
\gls*{sde}-\gls*{svae}  relied too heavily on the encoding network during training to reconstruct 
the state space which, despite the larger diffusion value, is reasonably disentangled. However, the 
drift function was not properly learned, which results in an overestimation of the diffusion and 
poor synthetic samples when the \gls*{sde}-\gls*{svae} cannot rely on probabilistic guesses of
the encoding network.

\subsection{Stochastic Lorenz system \label{sec:vaele:lorenz}}
\begin{figure}
    \centering
    \begin{subfigure}[t]{0.5\textwidth}
        \centering
        \includegraphics[width=\textwidth]{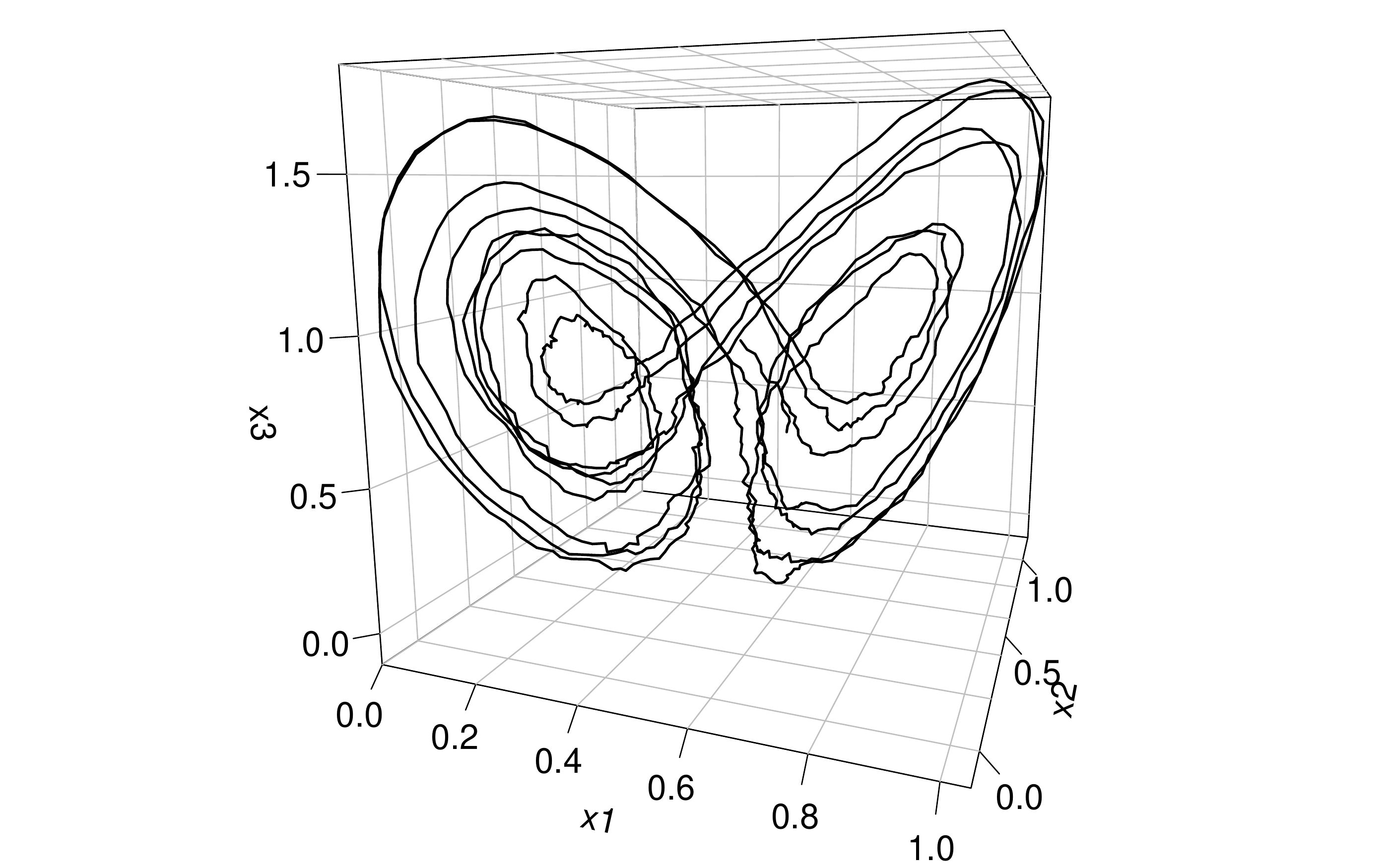}
        \caption{\label{fig:vaele:real_lorenz}}
    \end{subfigure}%
    \begin{subfigure}[t]{0.5\textwidth}
        \centering
        \includegraphics[width=\textwidth]{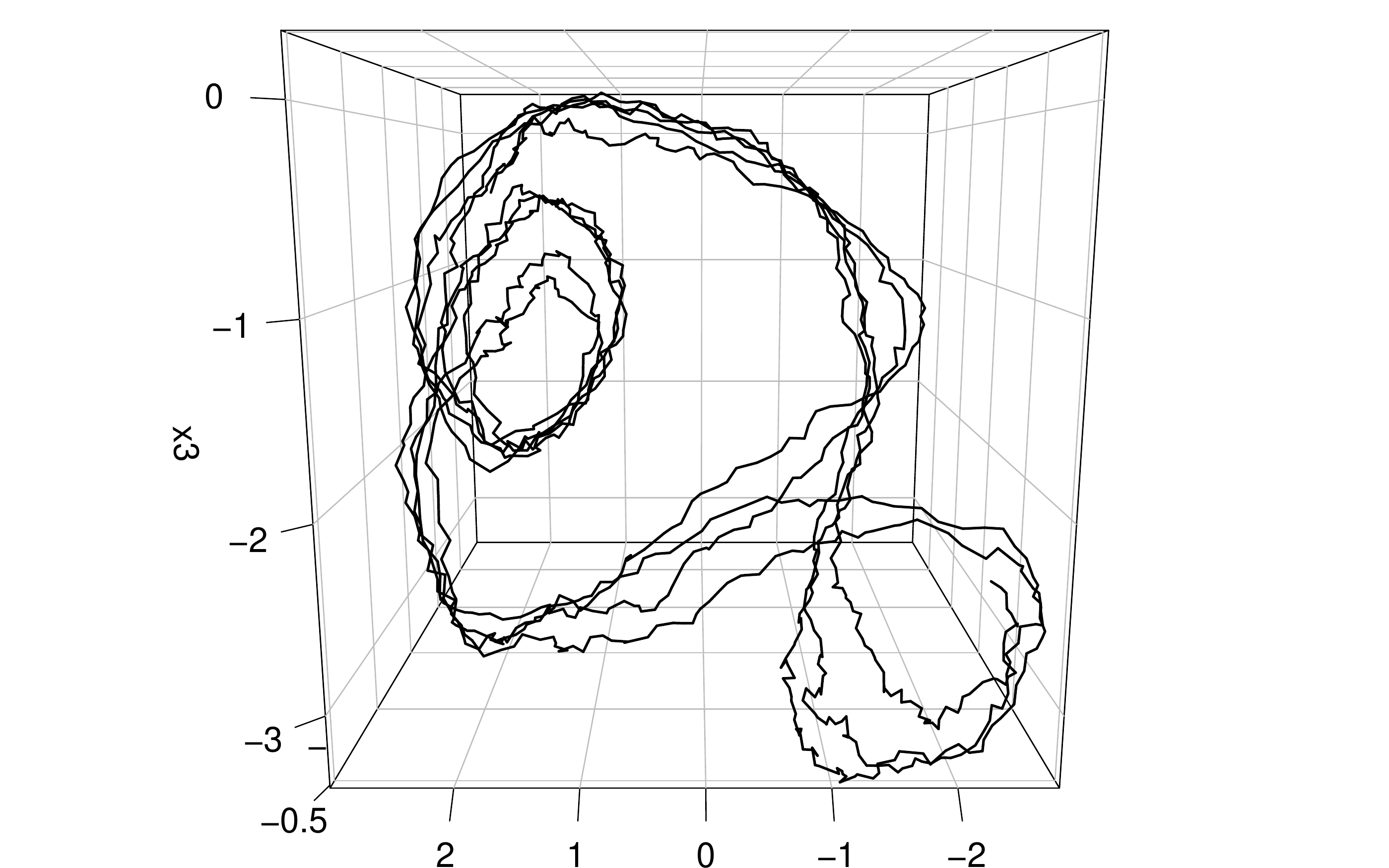}
        \caption{\label{fig:vaele:recons_lorenz}}
    \end{subfigure}
    \caption{
      Comparison of the (a) true stochastic Lorenz state space and (b) a reconstructed phase space after
      1000 iterations and the observation of experimental data.
    \label{fig:vaele:lorenz_ps}}
\end{figure}

In this section, we study a stochastic version of the Lorenz system, 
a simplified model for the study of atmospheric convection:
\begin{equation}
\begin{split}
    dx_1 &= \big(\sigma (x_2 - x_1)\big)dt  + dW_1(t), \\
    dx_2 &= \big(x_1 (\rho - x_3) - x_2\big)dt + dW_2(t), \\
    dx_3 &= (x_1 x_2 - \beta x_3)dt + dW_3(t).
    \label{eq:vaele:lorenz}
\end{split}
\end{equation}
We focused on the values $\sigma = 10, \beta =8/3, \rho =28$ which, in the 
case of the deterministic model, are known to produce chaotic behavior. A
representative portion of the real state space of the system is shown in 
Figure~\ref{fig:vaele:real_lorenz}.  The \gls*{svae} was fed with 200 different 
realizations of the $x_1(t)$ signal, sampled at 1000 Hz. Each of the resulting 
input signals, $y^{(r)}_{t} = x^{(r)}_1(t / 1000)$ had a length of $200$ samples.
We used $d=3$ for the state space dimension, $N=200$ for the size of the graphical
model, $S=10$ for the number of Monte Carlo samples, and $m=64$ for the number
of inducing points. The inducing points were initialized at the beginning of training 
using a grid-based approach in the range $[-3, 3]$.

\begin{figure}
  \begin{subfigure}[t]{0.5\textwidth}
    \centering
    \includegraphics[width=\textwidth]{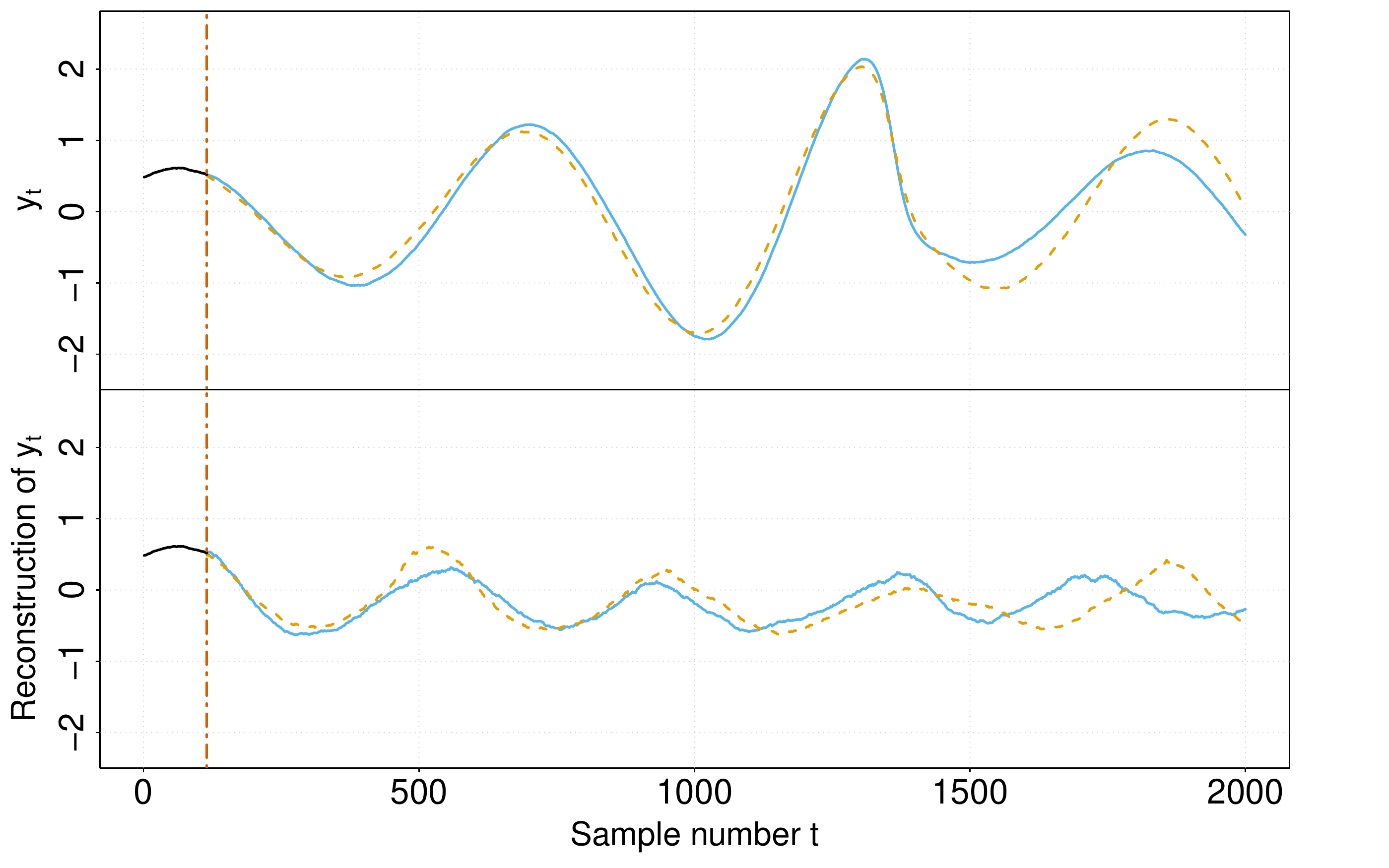}
    \caption{\label{fig:vaele:rossler_reconstruction}}
  \end{subfigure}
  \begin{subfigure}[t]{0.5\textwidth}
    \centering
    \includegraphics[width=\textwidth]{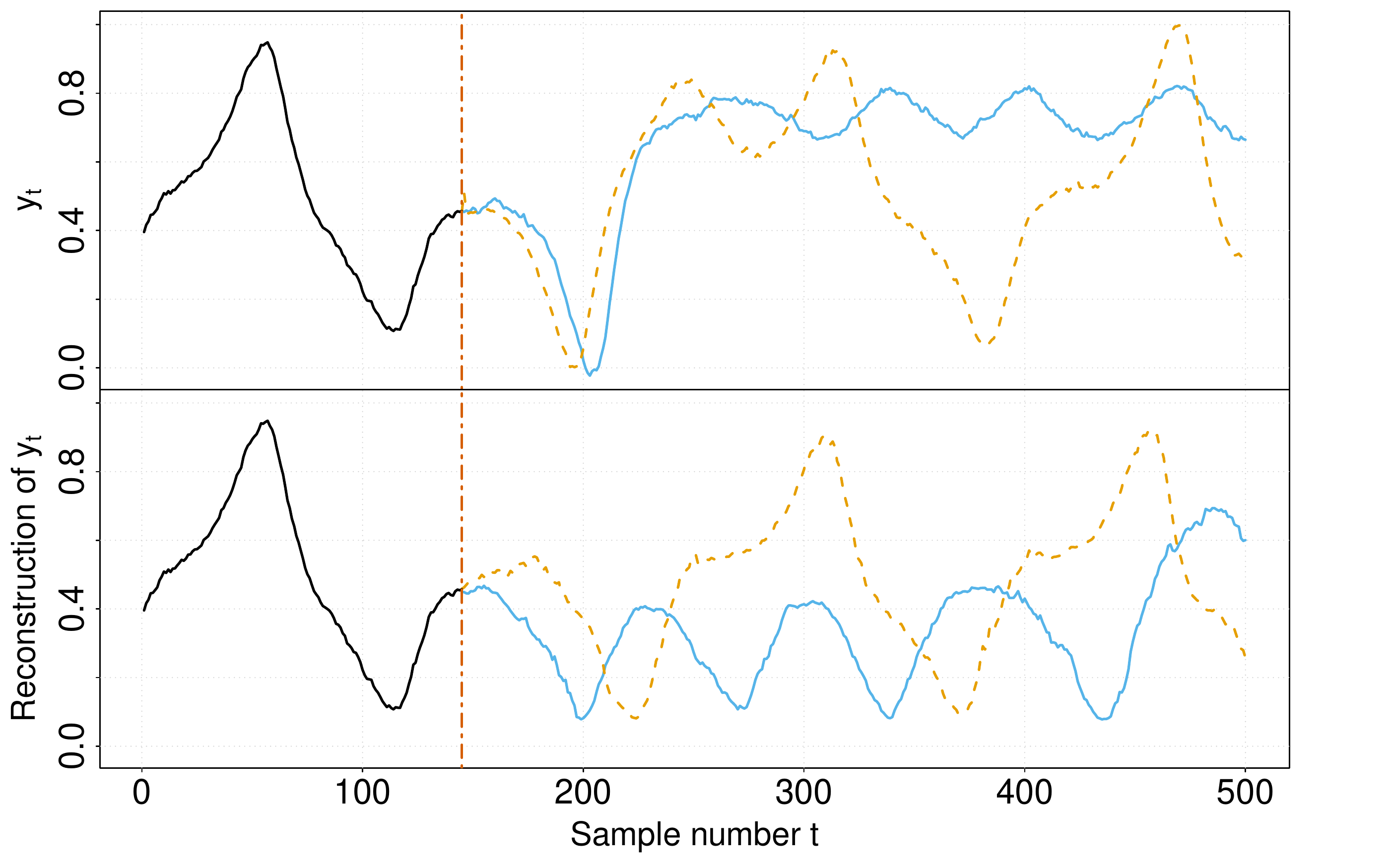}
    \caption{\label{fig:vaele:lorenz_prediction}}
  \end{subfigure}
  \caption{
  (a) Predictions from a \gls*{sde}-\gls*{svae} for the Rossler model.   
  (b) Predictions from a \gls*{sde}-\gls*{svae} for the Lorenz model.
  The panel and symbols used in the Figure are the same as in Figure~\ref{fig:vaele:reconstructions_1}.}
\end{figure}

Figures~\ref{fig:vaele:real_lorenz} and~\ref{fig:vaele:recons_lorenz} 
compare the real and a synthetic 3D-state space whereas that
Figure~\ref{fig:vaele:lorenz_prediction} shows predictions
generated by a trained \gls*{sde}-\gls*{svae}. In Figure~\ref{fig:vaele:recons_lorenz}, 
it is possible to identify the two small elliptical 
trajectories (at the top-left and bottom right of the figure) as each of the lobes
of the Lorenz attractor. The trajectories connecting both these regions represent 
transitions from one lobe to the other. Regarding Figure~\ref{fig:vaele:lorenz_prediction},
the synthetic new series resemble the original ones. However, lobe transitions probabilities 
differ in the original and synthetic series. As an illustrative example, synthetic time series
with sustained oscillations around $0.7$ (like the original blue one) were not observed. All 
synthetic time series performed only a few oscillations around $0.7$ before quickly transitioning
to the other lobe (see the dashed-orange example in the lower panel of Figure~\ref{fig:vaele:rossler_reconstruction}).

\section{Discussion and Conclusions\label{sec:vaele:discussion}}
In this paper, we have introduced a 
method that permits simultaneously learning a Markovian representation of the data 
and its dynamics, which are assumed to be described by a \gls*{sde} and hence represent
the state space of the system. Therefore, our method enables to bridge the gap between 
experimental time series and the use of \glspl*{sde} for modeling.

To this end, the \gls*{sde}-\gls*{svae} uses a Bayesian approach that leverages 
graphical models, which impose the restriction of having \gls*{sde}-like dynamics; 
and deep learning methods, which encodes probabilistic guesses which help in 
learning. Furthermore, \gls*{svae} takes advantage of \gls*{svi}
and natural gradients to implement an efficient inference algorithm, which permits 
the application of the method to large datasets that do not fit in main memory.
From our point of view, the main contribution of this work is exploiting the 
great ability of machine learning to detect data patterns through a model that 
remains interpretable within a prominent paradigm of statistical physics. Indeed, the
disentanglement of the dynamics in state space is achieved by only requiring 
that 1) similar state space vectors yield similar observed states, and 2) similar 
state space vectors result in similar dynamical evolution, described by a \gls*{sde}. 


The ability of the \gls*{sde}-\gls*{svae} to capture the main dynamical features
of the data and summarize them in the state space was shown using 
synthetic data. Although the results suggested that the \gls*{sde}-\gls*{svae} is
a powerful modeling tool, some drawbacks and future lines of work were also revealed.

A minor drawback of the method, which is shared with any technique using kernels, 
is that it requires a careful selection of a \gls*{gp} kernel  that it is able to
capture the relevant dynamics of the underlying process 
without loosing generalization capabilities. This is not always possible. For example, 
consider the use of the moment matching smoothing technique, which requires the use 
of squared exponential kernels. Although they may succeed in explaining the training 
data, the squared exponential kernels are only able to extrapolate around
``small'' regions around the inducing-points, where ``small'' is defined 
in terms of the length-scale hyperparameter. Hence, if a new experimental time series 
falls in a previously unexplored region of the state space, our 
technique will fail to predict the evolution of the system. It follows that, to improve the 
generalization properties of the drift estimates, we should endow 
the \gls*{sde}-\gls*{svae} with the ability of using different kernels depending on the 
dynamical properties of the system under study. Indeed, sophisticated kernels are usually 
hand-crafted to enable pattern discovery and extrapolation in specialized 
applications \cite{wilson2013gaussian}.  

An interesting difficulty when training the \gls*{sde}-\gls*{svae} is that
the autoencoder is trying to reconstruct a signal that is fed as input and that, unlike
most autoencoders, the latent space is not a bottleneck since its dimension is larger than 
the input dimension. Luckily, the variational formulation does impose constraints on 
the latent space, difficulting overfitting. However, as shown in Section~\ref{sec:vaele:rossler}
with the Rossler model, overfitting is still possible even though we tried to use small encoder networks for preventing 
the network from memorizing the input data. A reformulation of the lower bound may be needed 
to permit the use of larger and deeper networks while, at the same time, completely avoid
overfitting. For example, instead of reproducing the input time series, we may ask the \gls*{sde}-\gls*{svae}
to produce predictions several time steps ahead in the future. This may prevent overfitting 
and also places more emphasis in capturing the dynamics of the system.

Regarding the dynamics of the system, the experiments from Sections~\ref{sec:vaele:lotka} 
and~\ref{sec:vaele:lorenz} also show that the \gls*{sde}-\gls*{svae} does not fully 
reproduce all the relevant properties of the original time series. In the case of the 
Lotka-Volterra system, this is probably due to the fact that the system has a diffusion 
which is not compatible with the Lamperti transformation, which would permit its representation 
through \glspl*{sde} with constant diffusions, as assumed by our model. Hence, the use 
of non-constant diffusions should be considered in future extensions of the model. This may 
also help in better capturing the properties of other systems. According
to~\cite{brunton2017chaos}, deterministic dynamical systems with chaotic behaviour (such as the Lorenz system)
may be accurately modeled as a linear system with a forcing term. In most regions of the state 
space the forcing is small. When this term is not negligible, it acts as a rare-event forcing 
that drives lobe switching and approximates the nonlinear dynamics of the system. This forcing 
term could be naturally incorporated into a non-constant diffusion of a \gls*{sde}, which may 
help in capturing the properties of the system under study.

Another issue of the \gls*{sde}-\gls*{svae} is the computational time it 
requires to build the state space 
 when compared with simpler methods, like those based on lagged embeddings. While 
the latter are almost immediate, the \gls*{sde}-\gls*{svae} may need hundreds or even 
thousands of iterations until convergence (just like any method involving neural networks).
To alleviate this issue, an interesting line of 
work would be exploring initialization schemes that connect the efficient computation of 
lagged embeddings or Hankel-based embeddings with our method. Although in our experiments 
we fed the encoding network with lagged-embeddings, more sophisticated initialization schemes 
may accelerate convergence, facilitating the use of the \gls*{sde}-\gls*{svae} in the analysis 
of experimental data. Note however that an advantage of the \gls*{sde}-\gls*{svae} algorithm is that 
it is prepared for dealing with large datasets, whereas that most of the methods based on 
SVD are not.

Finally, it must be noted that the model proposed in this paper has 
room for improvement in light of the new advances in the field of machine learning. 
For example, \cite{NIPS2018_7892, li2020scalable} recently introduced continuous-time 
latent variable models, which is the natural way of formulating physical systems and 
also permits handling irregularly sampled time series. Also, our model is based on the
use of a variational posterior that factorizes the latent states ($\bm{x}$) and the 
drift function ($f(\cdot)$). More recently, several works started exploring the use 
of non-factorized posteriors~\cite{ialongo2019overcoming,curi2020structured,doerr2018probabilistic},
which may help in finding better embeddings and overcoming some of the issues found in
the experimentation. 

We believe that this work takes a valuable step towards developing an intuitive framework 
for building interpretable physical embeddings. To facilitate future research in this 
direction, we provide an open-source implementation of the method presented in this paper 
in \cite{github_vaele}.

\bibliography{references}  

\appendix
\section{Closed form natural gradients\label{app:nat_grad}}
Let us consider the distribution of the inducing points $\bm{\tilde{f}}$ and the 
precision parameter $\lambda$ (inverse of the diffusion) of the 
\gls*{sde} describing the evolution of the $i$-th component of the state space. 
According to Equation~\eqref{eq:vaele:full_model} the joint distribution of a finite set of drift
points and the precision parameter is a Gaussian-Gamma distribution. The 
Gaussian-Gamma distribution can be written in the natural form as
\begin{equation}
    \begin{split}
      \log q(\bm{\tilde{f}}, \lambda, \mid \alpha, \beta, \bm{\mu}, \bm{\Sigma}) =& 
     \begin{bmatrix} 
        \alpha-\frac12 \\
        -\beta-\frac{\bm{\mu}^T \bm{\Sigma}^{-1} \bm{\mu}}{2} \\
        \bm{\Sigma}^{-1} \bm{\mu}\\
        -\text{Vec}\big(\frac{\bm{\Sigma}^{-1}}{2}\big)\\
     \end{bmatrix} ^ T
     \begin{bmatrix} 
        \log \lambda\\
        \lambda \\
        \lambda \bm{\tilde{f}} \\
        \lambda \text{Vec}(\bm{\tilde{f}} \bm{\tilde{f}}^T)
     \end{bmatrix} + \\
     &- \bigg(
         \log \Gamma(\alpha)
         -\alpha \log \beta 
         - \frac12 \log \mid \bm{\Sigma}^{-1} \mid
     \bigg),
 \end{split}
\end{equation}
or, in terms of the natural parameters
$$\bm{\eta} = \Big[\eta_0, \eta_1, \bm{\eta}_2, \bm{\eta}_3 \Big]^T =
\Bigg[\alpha - \frac{1}{2},-\beta-\frac{\bm{\mu}^T \bm{\Sigma} ^{-1} \bm{\mu}}{2},
    \bm{\mu}^T\bm{\Sigma} ^{-1} ,
\text{Vec}\Bigg(-\frac{\bm{\Sigma}^{-1}}{2}\Bigg)^T
\Bigg]^T,$$
as
\begin{equation}
    \begin{split}
      \log q(\bm{\tilde{f}}, \lambda, \mid \bm{\eta}) \propto &
     \begin{bmatrix} 
        \eta_0 \\
        \eta_1\\ 
        \bm{\eta}_2\\
        \bm{\eta}_3\\ 
     \end{bmatrix} ^ T
     \begin{bmatrix} 
        \log \lambda\\
        \lambda \\
        \lambda \bm{\tilde{f}} \\
        \lambda \text{Vec}(\bm{\tilde{f}} \bm{\tilde{f}}^T)
     \end{bmatrix} + \\
     &- \bigg(
         \log \Gamma(\eta_0 + 1/2)
         -\big(\eta_0 + 1/2) \log \left(-\eta_1 +\frac{\bm{\eta}_2^T\text{Mat}(\bm{\eta}_3)^{-1}\bm{\eta_2}}{4} \right)
         - \frac12 \log \mid \text{Mat}(-2\bm{\eta_3}) \mid
     \bigg) \\
    =& \begin{bmatrix} 
       \eta_0 \\
       \eta_1\\ 
       \bm{\eta}_2\\
       \bm{\eta}_3\\ 
     \end{bmatrix} ^ T
     \begin{bmatrix} 
       T_0(\tilde{\bm{f}}, \lambda)\\
       T_1(\tilde{\bm{f}}, \lambda)\\
       T_2(\tilde{\bm{f}}, \lambda)\\
       T_3(\tilde{\bm{f}}, \lambda)\\
     \end{bmatrix}
     - A(\bm{\eta}) 
      = \bm{\eta}^T \bm{T}(\tilde{\bm{f}}, \lambda) - A(\bm{\eta}),
     \label{eq:natural_distro}
 \end{split}
\end{equation}
where $\text{Mat}(\cdot)$ is the inverse of $\text{Vec}(\cdot)$ (i.e., it transforms a vector in a symmetric matrix),
and $\bm{T}(\tilde{\bm{f}}, \lambda)$ and $A(\bm{\eta})$ are the sufficient statistics
and the log-partition of the Normal-Gamma distribution.

Our aim is to write now the lower bound from Equation~\eqref{eq:vaele:lower_bound_2} in terms 
of the natural parameters $\bm{\eta}$, also assuming that the prior  can be written in terms
of the natural parameters $\bm{\hat{\eta}}$, i.e.\, 
$p(\bm{\tilde{f}}, \lambda, \mid \bm{\hat{\eta}})$. To that end, it is useful the following 
property of the distributions with natural parameters~\cite{dasgupta2011exponential}:
\begin{equation}
\mathbb{E}\left[T_k(\bm{\tilde{f}}, \lambda)\right] = \nabla_kA(\bm{\eta}),
\label{eq:exp_property}
\end{equation}
where $\nabla_k A(\bm{\eta})$ is a shortcut for $\nabla_{\bm{\eta}_k} A(\bm{\eta}).$ In 
the case of the Normal-Gamma distribution Equation~\eqref{eq:exp_property} implies 
\begin{equation}
  \begin{split}
    \mathbb{E}\left[T_0(\bm{\tilde{f}}, \lambda)\right] &=
    \mathbb{E}\left[\log \lambda\right] =
      \psi(\alpha) - \log \beta=
    \nabla_0 A(\bm{\eta}),\\
    \mathbb{E}\left[T_1(\bm{\tilde{f}}, \lambda)\right] &= 
    \mathbb{E}\left[\lambda\right] =
      \alpha/ \beta=
    \nabla_1 A(\bm{\eta}),\\
    \mathbb{E}\left[T_2(\bm{\tilde{f}}, \lambda)\right] &= 
    \mathbb{E}\left[\lambda\bm{\tilde{f}}\right] =
    \frac{\alpha}{\beta} \bm{\mu}=
    \nabla_2 A(\bm{\eta}),\\
    \mathbb{E}\left[T_3(\bm{\tilde{f}}, \lambda)\right] &= 
    \mathbb{E}\left[\lambda\bm{\tilde{f}}\bm{\tilde{f}}^T\right] =
    \frac{\alpha}{\beta} \bm{\mu}\bm{\mu}^T + \bm{\Sigma}=
    \text{Mat}\left(\nabla_3 A(\bm{\eta})\right).
  \end{split}
  \label{eq:exp_identities}
\end{equation}
Using Equation~\eqref{eq:exp_identities} in Equations~\eqref{eq:vaele:lower_bound_2} and~\eqref{eq:vaele:second_term}, 
and taking into account that $\bm{\tilde{f}}$ and $\lambda$ refer to a single dimension (the $i$-th dimension) of the \gls*{sde} results in
\begin{align}
  \mathcal{L}(\bm{\eta}) =&
  \sum_{r=1}^R \Bigg[
    \sum_{t=1}^{N-1} \log \mathcal{N}\left(
    x^{(r)}_{i,t+1}\mid x^{(r)}_{i, t} + \bm{A}_i^{(r)}\bm{\bm{\mu}}_i, \beta_i/\alpha_i\right)\\
                         &-\frac{1}{2}\sum_{t=1}^{N-1} \left[\bm{P}_i(\bm{x}^{(r)}_{t})+
                         \text{tr}\left(\bm{A}_i^{(r)}\bm{\Sigma}_i\left[\bm{A}_i^{(r)}\right]^T\right) \right]
                         + \frac{N-1}{2} \left[\psi^0(\alpha_i) - \log (\alpha_i)\right]\Bigg]\\
      &-\mathcal{KL}\left[\bm{\eta} \mid \bm{\hat{\eta}}\right] + \text{constant}\\
  =&\left[\nabla_0 A(\bm{\eta})\right]^T \frac{N-1}{2}
    -\frac12\left[\nabla_1 A(\bm{\eta})\right]^T\sum_{r=1}^R\left[\Delta \bm{x}_i^{(r)}\right]^T\Delta \bm{x}_i^{(r)}
+ \left[\nabla_2 A(\bm{\eta})\right]^T \sum_{r=1}^R \left[\bm{A}_i^{(r)}\right]^T \Delta \bm{x}_i^{(r)} \\
   &-\frac12 \left[\nabla_3 A(\bm{\eta})\right]^T \text{Vec}\left(\sum_{r=1}^R
   \left[\bm{A}^{(r)}_i\right]^T \bm{A}_i^{(r)}\right)
   - \mathcal{KL}\left[\bm{\eta} \mid \bm{\hat{\eta}}\right]\\
  =&\left[\nabla_{\bm{\eta}}A(\bm{\eta})\right]^T
\left[
  (N-1)/2,
  -\sum_{r=1}^R\left[\Delta \bm{x}^{(r)}_i\right]^T\Delta \bm{x}_i^{(r)}/2,
  \sum_{r=1}^R \left[\Delta \bm{x}^{(r)}_i\right]^T\bm{A}_i^{(r)} ,
  -\text{Vec}\left(\sum_{r=1}^R \left[\bm{A}^{(r)}_i\right]^T \bm{A}_i^{(r)}\right)^T /2
\right]^T \\
   &-\mathcal{KL}\left[\bm{\eta} \mid \bm{\hat{\eta}} \right], \label{eq:lower_bound_on_eta}
\end{align}
where we have noted, to keep the notation uncluttered, $\Delta \bm{x}_i^{(r)} = \Delta \bm{x}^{(r)}_{i, 1:(N-1)}$,
$\bm{A}^{(r)}_i = \bm{A}_i\left(\bm{x}^{(r)}_{1:(N-1)}\right)$, and
$\mathcal{KL}\left[\bm{\eta} \mid \bm{\hat{\eta}}\right] =
\mathcal{KL}\left[
  q(\bm{\tilde{f}}, \bm{\lambda}\mid \bm{\eta}, \bm{\theta} ^ *)
  \mid p(\bm{\tilde{f}}, \bm{\lambda}\mid \bm{\hat{\eta}}, \bm{\theta} ^ *)
\right].$

Given that the Kullback-Leibler divergence between two distributions with natural parameters can be 
written as
$$
\mathcal{KL}\left[\bm{\eta} \mid \bm{\hat{\eta}}\right] =
\left[\nabla_{\bm{\eta}} A(\bm{\eta})\right]^T\left(\bm{\eta}-\bm{\hat{\eta}}\right)
-A(\bm{\eta})
+A(\bm{\hat{\eta}}),
$$
taking gradients in Equation~\eqref{eq:lower_bound_on_eta} results in
\begin{align}
  \nabla_{\bm{\eta}}\mathcal{L}(\bm{\tilde{f}}, \lambda)
  =&
  \left[\nabla^2_{\bm{\eta}}A(\bm{\eta})\right]^T
  \left[
    (N-1)/2,
    -\sum_{r=1}^R\left[\Delta \bm{x}_i^{(r)}\right]^T\Delta \bm{x}_i^{(r)}/2,
  \sum_{r=1}^R \left[\Delta \bm{x}_i^{(r)}\right]^T\bm{A}_i^{(r)} ,
  -\text{Vec}\left(\sum_{r=1}^R \left[\bm{A}^{(r)}_i\right]^T \bm{A}_i^{(r)}\right)^T /2
  \right]^T \\
   &-
   \left[\nabla^2_{\bm{\eta}} A(\bm{\eta})\right]^T\left(\bm{\eta}-\bm{\hat{\eta}}\right). \label{eq:standard_grad}
\end{align}

Finally, and following~\cite{amari1998natural}, the natural gradients (noted with $\tilde{\nabla}$)
can be computed using
\begin{equation}
  \tilde{\nabla}\mathcal{L}(\bm{\eta}) \triangleq \bm{F}_{\bm{\eta}}^{-1}\nabla
  \mathcal{L}(\bm{\eta}),
\end{equation}
where $\bm{F}_{\bm{\eta}}$ is the Fisher information matrix. Furthermore, the Fisher information 
matrix of a member of the exponential family with natural parameters $\bm{\eta}$ is, according to~\cite{dasgupta2011exponential}, 
\begin{equation}
  \bm{F}(\bm{\tilde{\eta}}) = 
  \nabla^2_{\bm{\eta}}A(\bm{\eta}),
\end{equation}
which applied to Equation~\eqref{eq:standard_grad} implies that
\begin{equation}
  \tilde{\nabla}_{\bm{\eta}}\mathcal{L} =
  \left[
    (N-1)/2,
    -\sum_{r=1}^R\left[\Delta \bm{x}_i^{(r)}\right]^T\Delta \bm{x}_i^{(r)}/2,
  \sum_{r=1}^R \left[\Delta \bm{x}_i^{(r)}\right]^T\bm{A}_i^{(r)} ,
  -\text{Vec}\left(\sum_{r=1}^R \left[\bm{A}^{(r)}_i\right]^T \bm{A}_i^{(r)}\right)^T /2
    \right]^T     - \left(
    \bm{\eta} - \bm{\hat{\eta}}
  \right).
\end{equation}

\end{document}